\documentclass[twocolumn,twoside]{article}
\usepackage[utf8]{inputenc}
\let\OLDthebibliography\thebibliography
\renewcommand\thebibliography[1]{
  \OLDthebibliography{#1}
  \setlength{\parskip}{0pt}
  \setlength{\itemsep}{0pt plus 0.3ex}
}

\usepackage{graphicx}

\usepackage{geometry}
\usepackage{fancyhdr}
\usepackage{latexsym}
\usepackage[fleqn]{amsmath}
\usepackage{fancybox}
\usepackage{comment}

\newcommand{\dbar}{{d\mkern-7mu\mathchar'26\mkern-2mu}}

\title{The principles of adaptation in organisms and machines I: \\machine learning, information theory, and thermodynamics}
\author{Hideaki Shimazaki
}
\date{Graduate School of Informatics, Kyoto University}

\begin{document}

\twocolumn[
  \begin{@twocolumnfalse}
    \maketitle
    \begin{abstract}
How do organisms recognize their environment by acquiring knowledge about the world, and what actions do they take based on this knowledge? This article examines hypotheses about organisms' adaptation to the environment from machine learning, information-theoretic, and thermodynamic perspectives. We start with constructing a hierarchical model of the world as an internal model in the brain, and review standard machine learning methods to infer causes by approximately learning the model under the maximum likelihood principle. This in turn provides an overview of the free energy principle for an organism, a hypothesis to explain perception and action from the principle of least surprise. Treating this statistical learning as communication between the world and brain, learning is interpreted as a process to maximize information about the world. We investigate how the classical theories of perception such as the infomax principle relates to learning the hierarchical model. We then present an approach to the recognition and learning based on thermodynamics, showing that adaptation by causal learning results in the second law of thermodynamics whereas inference dynamics that fuses observation with prior knowledge forms a thermodynamic process. These provide a unified view on the adaptation of organisms to the environment.
\vspace{1em}
    \end{abstract}
  \end{@twocolumnfalse}
]

{
  \renewcommand{\thefootnote}%
    {\fnsymbol{footnote}}
  \footnotetext[1]{Address: Yoshida-honmachi, Sakyo-ku, Kyoto 606-8501, Japan; Email: h.shimazaki@i.kyoto-u.ac.jp; URL: http://www.neuralengine.org}
}

\section{Introduction}\label{sec:prologue}
How do organisms recognize their environment by acquiring knowledge about the outside world for their survival, and what actions do they take based on this knowledge? The purpose of this article is to gain a deeper understanding of this issue by explaining the perception and behavior of organisms from multiple perspectives in different disciplines, namely machine learning, information theory, statistical mechanics, and thermodynamics.

Natural stimuli have characteristic statistical regularity that follows the rules of materials or life. As a result of evolution and development, it is thought that the nervous systems of living things adapt to the statistical regularity of natural stimuli, and efficiently encode them in their activity \cite{attneave1954some,barlow1961possible,barlow1972single,field1987relations,linsker1988self,atick1992does}. From an information theoretic point of view, Horace Barlow proposed the efficient coding hypothesis, an adaptive principle of brains, which states that redundancy of information in the environment is eliminated and represented as independent activity in the brain \cite{barlow1961possible,barlow1972single}. According to the efficient coding hypothesis, an understanding of the statistical structure of natural stimuli such as the sounds of forests, coastal landscapes or flowing streams deepens our understanding of coding mechanisms of the brain \cite{laughlin1981simple,olshausen1996emergence}. Indeed, it is known that this hypothesis explains response characteristics of early sensory neurons such as nonlinearity of responses to stimuli, or sensitivity to the temporal and spatial arrangement of stimuli, called receptive fields \cite{laughlin1981simple,olshausen1996emergence,olshausen1997sparse,simoncelli2001natural,schwartz2001natural}.

Natural stimuli must originate from certain causes. If we further consider the adaptive mechanisms of the brain, we arrive at the concept of the brain as an organ that infers the causes of input stimuli. According to this view, the brain is an inference organ equipped with an empirical model of what sorts of things in the outside world cause the sensory data, and infers the causes of the current data based on this model. This view was proposed in the first half of the 20th century by a German physicist and physician, Hermann von Helmholtz \cite{helmholtz1925treatise}.\footnote{Building on earlier work by Francis Bacon, Thomas Hobbes and others.} He used the term unbewusster Schluss (unconscious conclusions) to refer to conclusions about the causes of sensory input, arguing that these are dominated by inductive conclusions originating from experiences and analogy rather than deductive logic.\footnote{Helmholtz himself concludes the chapter on recognition in his book \cite{helmholtz1962helmholtz} as follows: “it is the characteristic function of the intellect to form general conceptions, that is, to search for causes; and hence it can conceive (begreifen) of the world only as being causal connection.”} This concept is now known as Helmholtz's unconscious inference. Statistically optimal reasoning is achieved by constructing a posterior distribution of causes from a Bayesian formula together with a hierarchical model (generative model) that describes the statistical structure of data generation. The hypothesis that organisms achieve optimal or approximate Bayesian inference using their nervous systems is generally called the Bayesian brain hypothesis \cite{doya2007bayesian}. In fact, in tasks where decisions have to be made based on uncertain input stimuli, many studies have reported that humans and other animals arrive at conclusions that are close to those from the Bayesian statistical inference \cite{ma2006bayesian}. Many ideas have been proposed by theoretical groups regarding mechanisms for calculating Bayesian posterior distributions with neural systems \cite{ma2006bayesian,pouget2013probabilistic,beck2011marginalization,funamizu2016neural}. 

Hinton and Dayan et al. proposed an optimization algorithm for a hierarchical model called the Helmholtz machine, and introduced variational free energy as its objective function \cite{hinton1994autoencoders,dayan1995helmholtz}. Friston et al. argued that various related theories in recognition and learning can be handled in a unified way by the principle of minimizing variational free energy (free energy principle) \cite{friston2003learning,friston2006free,friston2010free,friston2012free,buckley2017free,bogacz2017tutorial}. This principle is also called surprise minimization because data is no longer surprising after learning. In the literature of machine learning, the same framework for stochastic latent variable models is called variational inference \cite{blei2017variational,zhang2017advances}. This framework originates from Dempster et al.’s Expectation Maximization algorithm, and is established as an optimization framework of applied models including various hidden Markov models, latent dirichlet allocation used in natural language processing, and a variational autoencoder used in deep learning.

These models construct a generative model of the external world, which is a hypothesis about data generation. Among them, predictive coding theory of the brain \cite{mumford1992computational,kawato1993forward,rao1999predictive,lee2003hierarchical,clark2013whatever} hypothesizes that recursive hierarchical modules learn the statistical regularity. These modules receive prediction from higher modules and send prediction error to the higher modules, which successfully explained contextual modulation of early visual neurons by the feedback prediction signals \cite{rao1999predictive}. An extended framework called message passing algorithms (belief propagation) on graphical models is discussed as computational architecture of the cortex \cite{lee2003hierarchical,pitkow2017inference}, in particular in the context of the free energy principle \cite{friston2008hierarchical,friston2010free,friston2012perceptions}. In this article, we will see how information about the data is absorbed into the model as prior knowledge through learning, and how this knowledge can act as a top-down modulation signal, from perspectives of machine learning, information-theory, and thermodynamics.

\begin{figure*}[t]
\begin{center}
\includegraphics[width=.9\textwidth]{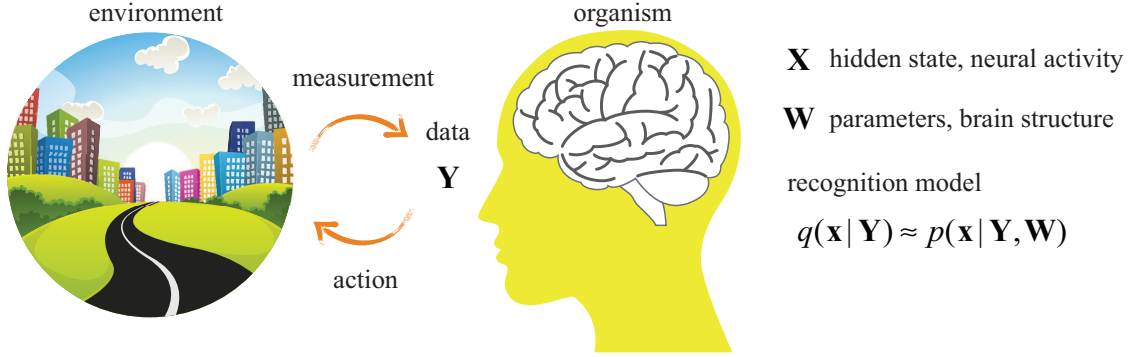}
\end{center}
\caption{Interaction between organisms and the environment}
\label{fig:brain_env}
\end{figure*}

These modeling approaches, however, take into account only passive aspects of the brain responding to presented stimuli. In real world, animals actively explore the environment to acquire data of the outside world. Indeed, as has been pointed out by many scientists and philosophers including Merleau-Ponty and, more recently, Andy Clark, Rodney Brooks and Francisco Valera, the brain is first and foremost a device for moving the body, and it is not possible to conclude the theory of recognition and learning without considering organisms' active interaction with the environment \cite{clark1998being,varela2017embodied}. 

Friston’s theory builds on the static recognition model to provide a unified view that also includes actions \cite{friston2006free,friston2010action,friston2012active,friston2012perceptions}. This hypothesis called active inference assumes that actions are selected based on the principle of least surprise. For example, consider the task of recognizing the book you are holding. To recognize a book using the static hierarchical models (including deep neural networks), it is necessary to perform learning by using a large amount of data obtained by tilting the book, rotating it, examining the cover and so on, so that it is possible to recognize an item as a “book” no matter what state it is in. But how do humans do this? If you do not understand what something is by looking only at the spine of the book, you might try turning this object over. When you see the front, you will recognize that it is a book. In this example, the use of actions makes it easier to recognize that the object is a book, even when using an internal model that only points to the cover of the book. When data does not match our generative model of the world, it is more likely to surprise us (because the goodness-of-fit of the data to the model is small). We can therefore make changes to the outside world in order to adapt the object to our recognition capabilities and reduce the surprise. This is a way of interpreting behavior based on the generative model of the world in the brain, and attempts are being made to create a unified theory of recognition and behavior by rewriting actions in this framework, including the selection of actions by reinforcement learning \cite{friston2009reinforcement,friston2012active,schwartenbeck2013exploration}. 

In this article, we first introduce the hierarchical model of the environment, and explain how the model is interpreted as a model of the brain. We review an approximate inference method in machine learning, and explain how it is related to the free energy principle. Next, we present an overview of learning hierarchical models from an information theoretic viewpoint, and explain how it relates to the information maximization principle (infomax principle) and the efficient coding hypothesis, which are the classical theories of sensory perception. Finally, from the framework of statistical mechanics and thermodynamics for entropy maximization, we redefine free energy by revisiting the model of recognition, and discuss the formal relationship among statistical learning, minimization of free energy, and the second law of thermodynamics.

\section{Learning hierarchical models}\label{sec:learning}
In this section, we introduce the hierarchical models used in the fields of machine learning and statistics, and consider the meaning of introducing hierarchical models as a model of the brain. Next, we define statistical learning and clarify difficulties of learning the hierarchical models to motivate the need of approximate inference methods that will be described in detail in the next section.

\subsection{Hierarchical models}
Organisms and machines can infer underlying causes from the observed data, and can use this knowledge in performing their next actions (see Fig.~\ref{fig:brain_env}). It was hypothesized that organisms achieve these recognition and behaviors by constructing a model on causal relationships in the outside world.\footnote{Here the `cause' or `causative factor' is the one assumed in the model or brain. To verify if the assumed causes indeed causally influence data, we need to intervene the world to control the causes. It is an act to test the causal model of the world. We will touch upon action selection based on the hierarchical model in the next section.} Consider a model that assumes the following hierarchical causal relationship. Let ${\mathbf{Y}}$ be the data obtained from the outside world, and assume that this data is sampled from a distribution $\bar p({\mathbf{y}})$.\footnote{In this article, we will use uppercase letters to denote random variables, and lowercase letters to denote variables.} The symbol ${\mathbf{X}}$ represents the causative factors behind the generation of data. Furthermore, ${\mathbf{W}}$ represents the environmental factors or contexts that underlie these causative factors and data. Assuming the distribution for the causes and environmental factors, the joint distribution $p({\mathbf{y}},{\mathbf{x}},{\mathbf{w}})$ that represents the dependency between the data and these factors is called a generative model. The following hierarchical model is now constructed as a generative model for data:
\begin{equation}
p({\mathbf{y}},{\mathbf{x}},{\mathbf{w}})=p({\mathbf{y}} \vert {\mathbf{x}}, {\mathbf{w}})p({\mathbf{x}} \vert {\mathbf{w}}) p({\mathbf{w}}).
\label{eq:generative_model1}
\end{equation}
In this article, the environmental factors ${\mathbf{w}}$ are treated as a parameter without assuming a distribution.\footnote{${\mathbf{x}}$ may be called the parameter of the observation model. In this case, ${\mathbf{w}}$ is called the hyper-parameter.} In this case, $p({\mathbf{y}}|{\mathbf{x}},{\mathbf{w}})$ is called the observation model, and $p({\mathbf{x}}|{\mathbf{w}})$ is called the prior distribution. A model of the data distribution can be derived by the following equation,
\begin{align}
p({\mathbf{y}}|{\mathbf{w}})
& = \int p({\mathbf{y}},{\mathbf{x}}|{\mathbf{w}}) d{\mathbf{x}}.
\label{eq:generative_model2}
\end{align}
This operation is called marginalization of latent variables.

A generative model is constructed based on an assumed hierarchical structure for the data generation, and this generative model can be used to estimate underlying causes from data according to the following Bayes' theorem: 
\begin{equation}
p({\mathbf{x}} |{\mathbf{Y}},{{\mathbf{w}} })
= \frac{{p({\mathbf{Y}}|{\mathbf{x}},{\mathbf{w}})p({\mathbf{x}} |{\mathbf{w}})}}{{p({\mathbf{Y}}|{{\mathbf{w}} })}}.
\label{eq:exact_posterior}
\end{equation}
This is the distribution of causative factors for a given set of data, and is called the posterior distribution. In the case of machine learning, estimation based on the posterior distribution may be performed by using its analytical formula, or by using sampling techniques. Organisms are also able to act based on inference and prediction, whereby sensory data is encoded in the form of neural activity and the network structure that supports it; therefore it is likely that the brain implements functions similar to the Bayesian inference \cite{doya2007bayesian}. Below we discuss how we consider the hierarchical model as a model of the brain in this article.


As a model of the brain, the posterior distribution of the latent variable $\mathbf{X}$ is interpreted as neural activity caused by the stimulus ${\mathbf{Y}}$. The prior distribution of $\mathbf{X}$ is neural activity exposed to the various stimuli in natural environment. To explain this in detail, we introduce the following simplified hierarchical structure as an internal model of the world in the brain:
\begin{equation}
p({\mathbf{y}},{\mathbf{x}} \vert {\mathbf{w}})=p({\mathbf{y}} \vert {\mathbf{x}}, {\boldsymbol{\phi}})p({\mathbf{x}} \vert {\boldsymbol{\lambda}}),
\label{eq:generative_model_of_brain}
\end{equation}
where ${\boldsymbol{\phi}}$ is a parameter of the observation model, ${\boldsymbol{\lambda}}$ is a parameter of the prior distribution, and ${\mathbf{w}}=({\boldsymbol{\phi}},{\boldsymbol{\lambda}})$.\footnote{From the original hierarchical model, this hierarchical model is obtained as follows. First, the observation model is obtained as $p({\mathbf{y}} \vert {\mathbf{x}}, {\boldsymbol{\phi}})$ under the assumption that ${\mathbf{y}}$ is conditionally independent of ${\boldsymbol{\lambda}}$ given ${\mathbf{x}}$. Second, the prior distribution is obtained as $p({\mathbf{x}} \vert {\boldsymbol{\lambda}})$ given that ${\mathbf{x}}$ does not depend on ${\boldsymbol{\phi}}$. Note that ${\mathbf{x}}$ is independent from ${\boldsymbol{\phi}}$ because the node ${\mathbf{y}}$ that connects these two is marginalized.}
The observation model $p({\mathbf{y}}|{\mathbf{x}},{\boldsymbol{\phi}})$ represents how data is expressed by combinations of neuronal activity ${\mathbf{X}}$. The parameter ${\boldsymbol{\phi}}$ can be the basis functions for representing the data. The prior distribution $p({\mathbf{x}}|{\boldsymbol{\lambda}})$ represents constraints on the neural activity.

\begin{table}[t]
  \begin{tabular}{|l | l | l } \hline
    Symbols   & Descriptions \\ \hline
    ${\mathbf{Y}}$, ${\mathbf{y}}$	& data \\
    & sensory stimulus \\
    ${\mathbf{X}}, {\mathbf{x}}$  & cause, latent variable \\
                            & neural activity\\
    ${\mathbf{W}}, {\mathbf{w}}$    & parameter, context \\
                            & brain structure \\
    ${\boldsymbol{\phi}}$ 	& parameter, basis function \\
    & brain structure\\
    ${\boldsymbol{\lambda}}$& parameter in the prior \\
                            & \\
    ${\boldsymbol{\omega}}$ & basis function, regularization \\
                            & structure of spontaneous activity \\
    ${\boldsymbol{\beta}}$  & regularization \\
                            & feedback modulation \\
    \hline \hline
    $p({\mathbf{Y}}|{\mathbf{w}})$  
                            & marginal likelihood, evidence \\
    &\\
    $p({\mathbf{Y}},{\mathbf{X}} |{\mathbf{w}})$ 
                            & complete data likelihood \\
    &\\
    $p({\mathbf{y}},{\mathbf{x}} |{\mathbf{w}})$
                            & generative model\\
    &\\
    $p({\mathbf{y}}|{\mathbf{x}},{\boldsymbol{\phi}})$ 
                            & observation model\\
    &\\
    $p({\mathbf{x}} |{\boldsymbol{\lambda}})$ 
                            & prior distribution\\
                            & spontaneous or modulation activity\\
    $p({\mathbf{x}} |{\mathbf{Y}},{\mathbf{w}})$
                            & posterior distribution\\
                            & evoked activity\\
    $q({\mathbf{x}}|{\mathbf{Y}})$
                            & approximate posterior distribution  \\
    					    & recognition model, evoked activity \\
     \hline \hline
     $\mathcal{L}[q,p]$ & lower bound, variational lower bound, \\
     					& evidence lower bound (ELBO) \\
     $\mathcal{Q}[q,p]$	& expected complete data\\
     					& log-likelihood, $\mathcal{Q}$-function\\
     $H[q]$				& entropy\\
     \hline
  \end{tabular}
\end{table}

Next, using this generative model, the posterior distribution of the latent variable is obtained as
\begin{equation}
p({\mathbf{x}} |{\mathbf{Y}},{{\mathbf{w}} })
= \frac{{p({\mathbf{Y}}|{\mathbf{x}},{\boldsymbol{\phi}})p({\mathbf{x}} | {\boldsymbol{\lambda}})}}{{p({\mathbf{Y}}|{{\mathbf{w}} })}}.
\label{eq:exact_posterior_brain}
\end{equation}
The posterior distribution of ${\mathbf{X}}$ represents the neural activity in response to stimulus ${\mathbf{Y}}$.\footnote{Alternatively, often in literature on predictive coding and the free energy principle, activity rates (firing rates) of neurons refers to maximum a posteriori (MAP) estimate of the posterior distribution as a proxy of the distribution. Here we take a view that neural activity represents a sample from the posterior distribution.} It is formed by combination of the observation model and prior distribution. Thus we can interpret that the posterior is constructed by modulating the internal neural activity represented as the prior distribution by observing data. In this case, we identify the prior as spontaneous activity of neurons \cite{berkes2011spontaneous}. Alternatively, we may consider the prior distribution constrains the neural activity; therefore act as a bias signal on the neural activity. For example, if ${\mathbf{X}}$ is assumed to be the activity of an area in the early visual cortex, the influence of the activity from other areas (such as feedback input from lateral neurons or upper areas) can be expressed by the prior distribution.

The above provides a view to interpret the neural activity as dynamics to achieve inference of the causative factors in the outside world. However, it should be noted that the flow of information in our physiological systems is opposite: external stimuli or causes make peripheral sensory receptors respond, which successively activate peripheral and central nervous systems. How this forward direction of the information processing can be framed into the inverse inference framework will be discussed in the section that relates the inference framework with information theory.  

\subsection{Learning}
Learning refers to the process of adjusting the model parameters so that the true data distribution and the model distribution become as \textit{close} as possible. Here it is necessary to define the concept of closeness for distributions. We use the Kullback-Leibler divergence, according to which the closeness of a distribution to the true data distribution is defined as follows:
\begin{align}
{\rm{KL}}  [\bar p({\mathbf{y}})  || & p({\mathbf{y}}|{\mathbf{w}})] 
\equiv \int {\bar p({\mathbf{y}})\log \frac{{\bar p({\mathbf{y}})}}{{p({\mathbf{y}}|{\mathbf{w}})}}} d{\mathbf{y}} \nonumber \\
&= {E_{{\mathbf{Y}}}}\log \bar p({\mathbf{Y}}) - {E_{{\mathbf{Y}}}}\log p({\mathbf{Y}}|{\mathbf{w}}).
\label{KL_divergence}
\end{align}
Throughout this article, $E_{{\mathbf{Y}}}$ represents the expectation by the true data distribution $\bar p({\mathbf{y}})$.\footnote{Note that this is not the expected value of the data generated by the model.} We want to reduce this KL divergence by modifying the model through learning, and this can be done by maximizing ${E_{{\mathbf{Y}}}}\log p({\mathbf{Y}}|{\mathbf{w}})$ in the second term. The true data distribution ${E_{{\mathbf{Y}}}}$ is unknown, but this term can be estimated by replacing it with the distribution of the observed data (empirical distribution). When an independent sample has been obtained, the empirical distribution can be written as $p({\mathbf{y}}) = n^{-1} \sum\nolimits_{i = 1}^n {\delta ({\mathbf{y}} - {{\mathbf{Y}}_i})}$. For the sake of simplicity, we use ${\mathbf{Y}}$ to represent either a set of all samples $\{{{\mathbf{Y}}_1},{{\mathbf{Y}}_2},\ldots,{{\mathbf{Y}}_n}\}$ or one sample ${\mathbf{Y}}_i$ ($i=1,\dots,n$) (see discussion below).

\begin{figure}[t]
\begin{center}
\includegraphics[width=.5\textwidth]{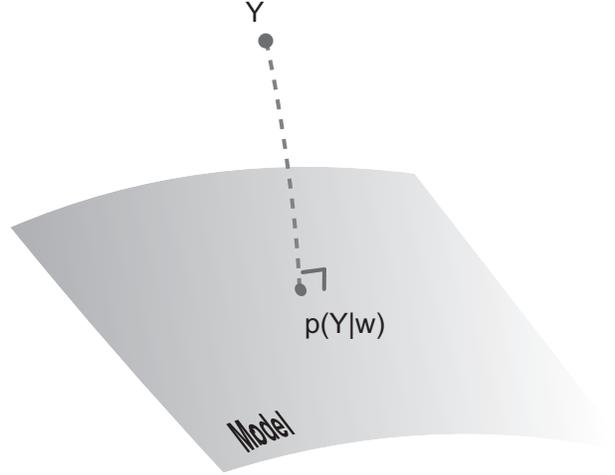}
\end{center}
\caption{Maximum likelihood estimation}
\label{fig:likelihood}
\end{figure}

Thus $\log p({\mathbf{Y}}|{\mathbf{w}})$ is the estimated value of the second term in Eq.~\ref{KL_divergence}, and is a function of ${\mathbf{w}}$ since the data point ${\mathbf{Y}}$ is given. $p({\mathbf{Y}}|{\mathbf{w}})$ is called a marginal likelihood function, and its logarithm is called the log marginal likelihood function. The process of determining parameters that maximize the marginal likelihood function is called the Type II maximum likelihood estimation. The resulting maximum likelihood estimate value is as follows:
\begin{equation}
{\mathbf{W}^{\ast}} = \arg {\max _{{\mathbf{w}}} } \log p({\mathbf{Y}}|{\mathbf{w}}).
\end{equation}
Here the marginal likelihood function is given by Eq.~\ref{eq:generative_model2}. The maximum likelihood estimation method is the process of choosing the model distribution closest to sample data in terms of the KL divergence (Fig.~\ref{fig:likelihood}). It is a projection of a sample from a higher-dimensional data distribution to a restricted model space.

The method for constructing a posterior distribution (Eq.~\ref{eq:exact_posterior}) using the generative model with the parameter ${\mathbf{W}^{\ast}}$ that is empirically obtained by learning (maximum likelihood estimation) is called the empirical Bayes method:
\begin{equation}
p({\mathbf{x}} |{\mathbf{Y}},{{\mathbf{W}}^{\ast} })
= \frac{{p({\mathbf{Y}}|{\mathbf{x}},{\mathbf{W}^{\ast}})p({\mathbf{x}} |{\mathbf{W}^{\ast}})}}{{p({\mathbf{Y}}|{\mathbf{W}^{\ast}})}}.
\label{eq:emprical_bayes}
\end{equation}
This estimation method is a hybrid of the Bayesian inference and maximum likelihood estimation. In this way, the empirical Bayes method that incorporates the “experience” of data-based learning into prior knowledge is an algorithm that realizes Helmholtz’s unconscious inference.

But unfortunately, this direct empirical Bayes method often runs into difficulties. To obtain the marginal likelihood of Eq.~\ref{eq:generative_model2}, it is necessary to integrate over the causative factors ${\mathbf{x}}$, but it is difficult to perform this integration analytically except when the observation model and prior distribution are normal distributions or special distributions. If the marginal likelihood and its gradient cannot be obtained, then it will not be possible to perform parameter optimization. Also, since the marginal likelihood appears as a normalization term in the Bayes' theorem of Eq.~\ref{eq:emprical_bayes}, it will also be impossible to calculate the posterior distribution. 

In statistics and machine learning, methods were therefore devised for approximately optimizing parameters and constructing the posterior distribution. The algorithm used for this approximate inference is described below. Since the inference processes in the brain’s neural network should be performed approximately, algorithms for learning approximate inference model in the statistics and machine learning is instructive when we consider models of the brain \cite{friston2003learning}. 

\begin{figure}[t]
\begin{center}
\includegraphics[width=.5\textwidth]{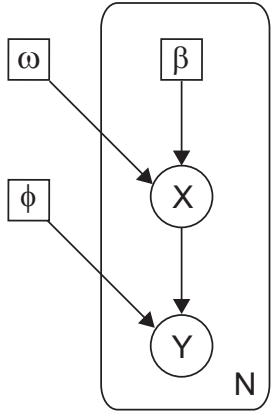}
\end{center}
\caption{Graphical representation of the hierarchical model.}
\label{fig:graphical_model}
\end{figure}

Finally, it is important to note that learning the parameters can be performed using different sets of data, namely either an entire set of data or each sample. Although expressions in the above and following descriptions do not distinguish these differences for the sake of simplicity, it should be noted that these different optimization strategies result in different learning dynamics that operate in distinct time-scales, and have significant implications of the parameters as a model of the brain. In this article, we assume that the parameters in the observation model ${\boldsymbol{\phi}}$ are learned from an entire set of the observed data, $\{{{\mathbf{Y}}_1},{{\mathbf{Y}}_2},\ldots,{{\mathbf{Y}}_n}\}$. The graphical representation of the model is shown in Fig.~\ref{fig:graphical_model}.  For example, the basis functions to represent images are learned from a set of many visual scenes exposed to an animal. In organisms, it is expected that learning with multiple samples are performed in an online manner that marginally updates parameters according to the contribution by each sample in turn. Thus this learning is gradual. The same approach, called the stochastic gradient descent (SGD) method, is taken in machine learning to deal with a large number of samples. 


In contrast, we can consider two types of parameters in the prior distribution ${\boldsymbol{\lambda}}=\{{\boldsymbol{\beta}},{\boldsymbol{\omega}}\}$: one set of parameters ${\boldsymbol{\omega}}$ are optimized by using all samples, and the others ${\boldsymbol{\beta}}$ are optimized for each sample of the data. For example, the parameters that represent spontaneous activity of neurons may be learned from many samples \cite{berkes2011spontaneous}, whereas the parameters that modulates representation by the neural activity according to the current context needs to be learned from each sample. This latter learning is faster than the others. In this case, learning the prior plays an important role in constructing a posterior distribution that is tailored for an individual sample. Early sensory neurons receive feedback/recurrent modulation via top-down or lateral connections, which plays an essential role in perceptual experiences \cite{super2001neural,Manita2015}, attention \cite{roelfsema1998object,reynolds2000attention}, and reward modulation \cite{Schultz2016} (see \cite{lamme2000distinct} for a review). This feedback/recurrent modulation is interpreted as a prior signal that biases the observation model, and construct the posterior (evoked activity of neurons). In machine learning, the same approach that optimize prior parameters for each sample is known as the automatic relevance determination (ARD) method \cite{bishop2006_pattern}, and is often used to represent the data by imposing sparsity to basis functions. 

\section{Approximate inference} \label{sec:approximate_inference}
In this section, we will explain the approximate inference methods that play an important role in machine learning, known as the Expectation-Maximization (EM) algorithm. This classical algorithm is a basis of variational methods which are standard tools to learn hierarchical models under the principle of minimizing variational free energy. Understanding the logic of fitting the model in the EM algorithm thus provides the basis of constructing advanced generative models, and is also instructive when we consider recognition and learning happening in the brain \cite{friston2003learning}. A detailed description of the EM algorithm can be found in standard machine learning textbooks as well \cite{mackay2003information,bishop2006_pattern}. 

\subsection{Recognition model and lower bound} \label{subsec:recognition_model}
Instead of using an exact posterior distribution, the approximate inference method considers another distribution that is easier to handle, and uses it to approximate the posterior distribution:
\begin{equation}
q({\mathbf{x}} |{\mathbf{Y}}) \approx p({\mathbf{x}} |{\mathbf{Y}},{\mathbf{w}}).
\end{equation}
This approximate posterior distribution is also referred to as a recognition model. Using this recognition model, the logarithmic marginal likelihood can be decomposed as follows \cite{kingma2017variational}: 
\begin{align}
&\log p({\mathbf{Y}}|{\mathbf{w}})  \nonumber\\
&\phantom{=}= \int {q({\mathbf{x}}|{\mathbf{Y}})\log p({\mathbf{Y}}|{\mathbf{w}})d{\mathbf{x}}  }  \nonumber\\
&\phantom{=}= \int {q({\mathbf{x}}|{\mathbf{Y}})\log \frac{{p({\mathbf{Y}},{\mathbf{x}}  |{\mathbf{w}})}}{{p({\mathbf{x}} |{\mathbf{Y}},{\mathbf{w}})}}d{\mathbf{x}} }  \nonumber\\
&\phantom{=}= \int {q({\mathbf{x}}|{\mathbf{Y}})\log \frac{{q({\mathbf{x}}|{\mathbf{Y}})}}{{q({\mathbf{x}}|{\mathbf{Y}})}}\frac{{p({\mathbf{Y}},{\mathbf{x}} |{\mathbf{w}})}}{{p({\mathbf{x}}  |{\mathbf{Y}},{\mathbf{w}})}}d{\mathbf{x}} }  \nonumber
\end{align}
\begin{align}
&\phantom{=}= 
 \underbrace{ \int {q({\mathbf{x}}|{\mathbf{Y}}) \log \frac{{p({\mathbf{Y}},{\mathbf{x}}  |{\mathbf{w}})}}{{q({\mathbf{x}} |{\mathbf{Y}})}}d{\mathbf{x}}  } }_{\text{Lower bound}}\nonumber\\
&\phantom{===} + \underbrace{ \int {q({\mathbf{x}}|{\mathbf{Y}})\log \frac{{q({\mathbf{x}}|{\mathbf{Y}})}}{{p({\mathbf{x}}  |{\mathbf{Y}},{\mathbf{w}})}}d{\mathbf{x}}  } }_{\text{KL divergence}}.
\end{align}
In the first equation, the term $\int {q({\mathbf{x}} |{\mathbf{Y}})d{\mathbf{x}} }  = 1$ is inserted. The second equation uses the relation ${p({\mathbf{Y}}|{\mathbf{w}})}= {p({\mathbf{Y}},{\mathbf{x}} |{\mathbf{w}})} / {{p({\mathbf{x}} |{\mathbf{Y}},{\mathbf{w}}}) }$ based on the formula of a posterior distribution (see Eq.~\ref{eq:exact_posterior}). In the third equation, a recognition model is inserted into the logarithm. 

Here, the first term obtained by the last equation, 
\begin{equation}
\mathcal{L}[q,p] \equiv \int {q({\mathbf{x}} |{\mathbf{Y}})\log \frac{{p({\mathbf{Y}},{\mathbf{x}} |{\mathbf{w}})}}{{q({\mathbf{x}} |{\mathbf{Y}})}}d{\mathbf{x}} },
\end{equation}
is called the evidence lower bound (ELBO) or variational lower bound, and plays a central role in this article. We will simply refer to it as the lower bound in this article. The negative value of the lower bound is called the variational free energy, which is the objective function in the free energy principle. In general, the lower bound $\mathcal{L}[q,p]$ is a functional of the recognition model $q({\mathbf{x}} |{\mathbf{Y}})$ and the generative model $p({\mathbf{Y}},{\mathbf{x}} |{\mathbf{w}})$. The second term is the KL divergence between the recognition model (an approximate posterior distribution) and the exact posterior distribution.\footnote{The expected value of this term according to $\bar p(\mathbf{y})$ is called the conditional KL divergence. Thus the second term is an estimate of the conditional KL divergence.} In summary, the log marginal likelihood is decomposed as follows:
\begin{equation}
\log p({\mathbf{Y}}|{\mathbf{w}}) = \mathcal{L}[q,p] + {\rm{KL}}[q({\mathbf{x}} |{\mathbf{Y}}) ||p({\mathbf{x}} | {\mathbf{Y}},{\mathbf{w}})].
\label{eq:loglikelihood_L_KL}
\end{equation}
Note that $\mathcal{L}[q,p]$ is called the lower bound because KL divergence takes a non-negative value, so the log marginal likelihood function is bound by this function.\footnote{The lower bound $\mathcal{L}[q,p]$ can also be obtained using Jensen’s inequality as:
\begin{align}
\log p({\mathbf{Y}}|{\mathbf{w}})
&= \log \int {p({\mathbf{Y}},{\mathbf{x}}|{\mathbf{w}})d{\mathbf{x}}}  \nonumber\\
&= \log \int {q({\mathbf{x}} |{\mathbf{Y}})\frac{{p({\mathbf{Y}},{\mathbf{x}} |{\mathbf{w}})}}{{q({\mathbf{x}} |{\mathbf{Y}})}}d{\mathbf{x}} }  \nonumber\\
&\geq \int {q({\mathbf{x}} |{\mathbf{Y}})\log \frac{{p({\mathbf{Y}},{\mathbf{x}} |{\mathbf{w}})}}{{q({\mathbf{x}} |{\mathbf{Y}})}}d{\mathbf{x}}} \nonumber\\
&\equiv \mathcal{L}[q,p]. \nonumber
\end{align}
}
\begin{equation}
\log p({\mathbf{Y}}|{\mathbf{w}}) \geq \mathcal{L}[q,p].
\end{equation}

The lower bound can be decomposed as follows:
\begin{align}
\mathcal{L}[q,p] &=   \underbrace{ \int {q({\mathbf{x}} |{\mathbf{Y}})\log p({\mathbf{Y}},{\mathbf{x}} |{\mathbf{w}})d{\mathbf{x}} }  }_{\text{$\mathcal{Q}$-function}}\nonumber\\
&\phantom{===} \underbrace{-  \int {q({\mathbf{x}} |{\mathbf{Y}})\log q({\mathbf{x}} |{\mathbf{Y}})d{\mathbf{x}} } }_{\text{Entropy}}.
\label{eq:lowerbound_decomposition}
\end{align}
Here, the first term is called the expected complete data log-likelihood function or $\mathcal{Q}$-function:
\begin{equation}
\mathcal{Q}[q,p] \equiv \int {q({\mathbf{x}} |{\mathbf{Y}})\log p({\mathbf{Y}},{\mathbf{x}} |{\mathbf{w}})d\theta }.
\end{equation}
The second term is the entropy of the recognition model.
\begin{equation}
H[q] \equiv  - \int {q({\mathbf{x}} |{\mathbf{Y}})\log q({\mathbf{x}} |{\mathbf{Y}})d{\mathbf{x}} }
\end{equation}
Using these equations, the EM algorithm alternates between optimizing the approximate posterior distribution and optimizing the parameters \cite{dempster1977maximum}. The actual steps of the EM algorithm are described below. 

\begin{figure*}[t]
\begin{center}
\includegraphics[width= \textwidth]{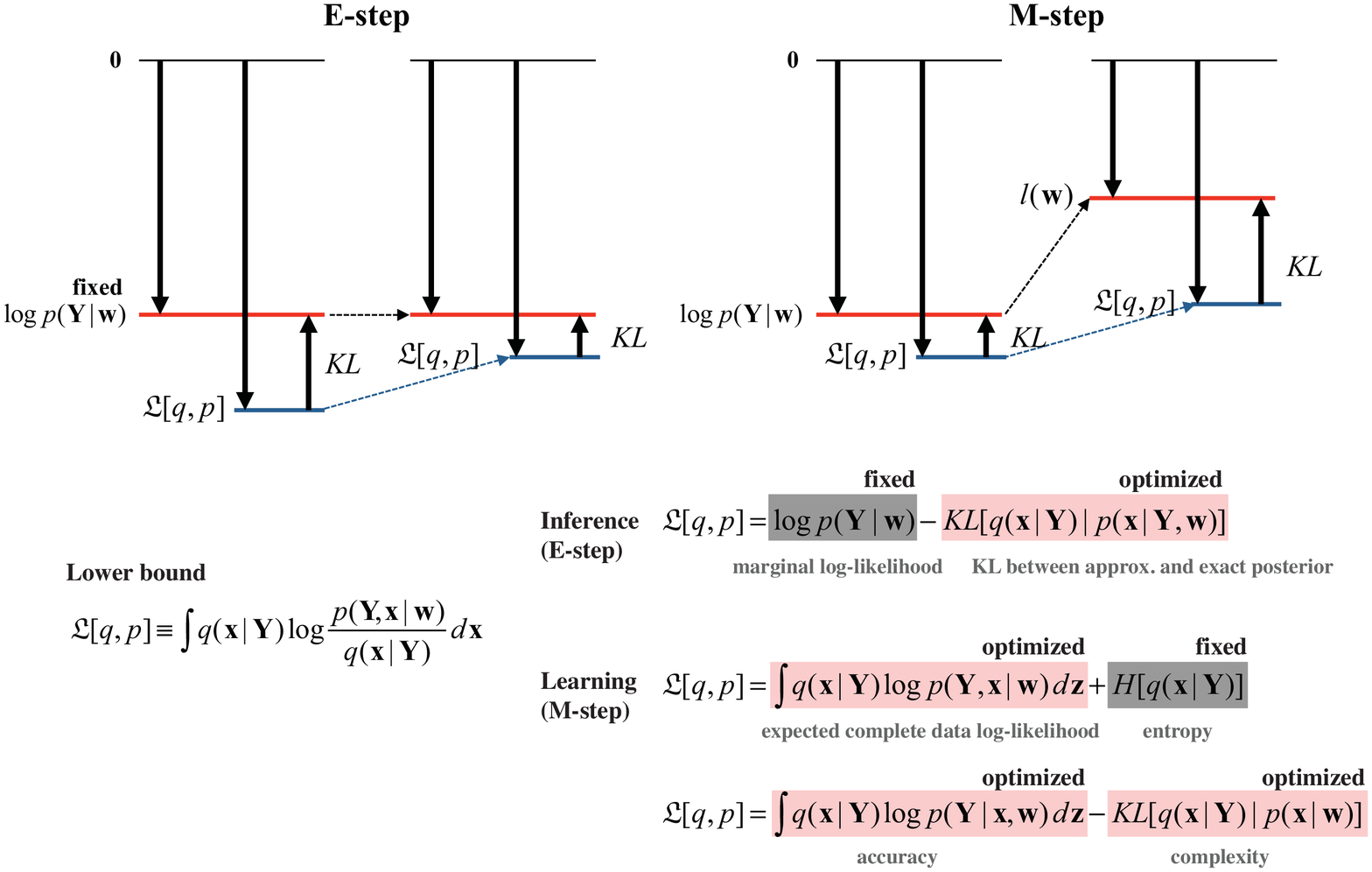}
\end{center}
\caption{The EM algorithm}
\label{fig:em_algorithm}
\end{figure*}

\subsection{EM algorithm} \label{subsec:em_algorithm}
The EM algorithm alternates between two steps called the E-step and the M-step. Figure \ref{fig:em_algorithm} shows a schematic representation of these steps.

\textbf{E-step}: \hskip .5em In the E-step, the recognition model is optimized to provide good approximation of the exact posterior distribution while the generative model is fixed. The purpose of this step is to ensure that the lower bound provides a tight bound on the marginal likelihood. First, note that the parameter ${\mathbf{w}}$ of the generative model is fixed, so the marginal likelihood $\log p({\mathbf{Y}}|{\mathbf{w}})$ is constant. In this case, according to Eq.~\ref{eq:loglikelihood_L_KL}, when KL divergence is reduced by approximating the recognition model to the exact posterior distribution, the lower bound becomes correspondingly larger and approaches the log marginal likelihood function.

Many methods have been proposed for approximating the posterior distributions, although this article will not go into the details of these methods. For example, stochastic methods include Monte Carlo methods, and deterministic methods include Laplace approximation (approximation by a Gaussian distribution), variational approximation (mean field approximation), and expectation propagation methods.

\textbf{M-step}: \hskip .5em The aim of the M-step is to increase the marginal likelihood by optimizing the parameter ${\mathbf{w}}$. In this step, the recognition model is fixed. It is presumed that optimization in the E-step results in the best approximation of the exact posterior distribution $p({\mathbf{x}} |{\mathbf{Y}},{{\mathbf{w}} }) $ within the range of the model assumed for the recognition model $q({\mathbf{x}} |{\mathbf{Y}})$, therefore the KL divergence of these two distributions (second term of Eq.~\ref{eq:loglikelihood_L_KL}) is minimized. In this case, the KL divergence is expected to increase due to the change in the exact posterior distribution caused by change of the parameter ${\mathbf{w}}$, which then contributes to increasing the marginal likelihood. Therefore, the change of parameters can be performed so as to increase the first term of Eq.~\ref{eq:loglikelihood_L_KL}, i.e., the lower bound. Incidentally, we have already seen that the lower bound can be decomposed as follows:
\begin{equation}
\mathcal{L}[q,p] = \mathcal{Q}[q,p] + H[q].
\end{equation}
Since the recognition model is now fixed, only the $\mathcal{Q}$-function of the first term is dependent on the parameters. Therefore, in the M-step, we use parameters that maximize the $\mathcal{Q}$-function.

In summary, when using E-step to optimize the approximate posterior distribution, the lower bound increases because the KL-divergence decreases with constant marginal likelihood. In M-step, the lower bound is increased by selecting parameters that maximize the $\mathcal{Q}$-function. Performing E-step and M-step alternately therefore causes the lower bound to increase monotonically, which is expected to increase the marginal likelihood. This justifies changing the objective function from the marginal likelihood to the lower bound $\mathcal{L}[q,p]$. As in the studies by Friston et al., we may construct a theory that starts from minimization of the variational free energy that is the negative lower bound. 

\subsection{Adaptation of the generative model and behavior}\label{subsec:lower_bound}
The above description shows how approximate inference is achieved by the EM algorithm. In the following, we will look into details of the learning step, considering cases where actions are included in this reasoning. By rearranging the expressions discussed above in terms of the lower bound $\mathcal{L}[q,p]$, we obtain the following equations for E-step and M-step, respectively:
\begin{align}
\mathcal{L}[q,p] &= \log p({\mathbf{Y}}|{\mathbf{w}}) - {\rm{KL}}[q({\mathbf{x}}|{\mathbf{Y}})|p({\mathbf{x}} |{\mathbf{Y}},{\mathbf{w}})], \nonumber\\
\mathcal{L}[q,p] &= \int{q({\mathbf{x}} |{\mathbf{Y}})\log p({\mathbf{Y}},{\mathbf{x}} |{\mathbf{w}}) } d{\mathbf{x}} + H[q]. 
\label{eq:lowerbound_qent_qfunc}
\end{align}
Furthermore, the lower bound can be decomposed as follows:
\begin{align}
\mathcal{L}[q,p]
&= \int {q({\mathbf{x}} |{\mathbf{Y}})\log \frac{{p({\mathbf{Y}},{\mathbf{x}} |{\mathbf{w}})}}{{q({\mathbf{x}} |{\mathbf{Y}})}}d{\mathbf{x}} } \nonumber\\
&= \int {q({\mathbf{x}} |{\mathbf{Y}})\log \frac{{p({\mathbf{Y}}|{\mathbf{x}},{\mathbf{w}})p({\mathbf{x}} |{\mathbf{w}})}}{{q({\mathbf{x}} |{\mathbf{Y}})}}d{\mathbf{x}} } \nonumber\\
%
\phantom{\mathcal{L}[q,p]}
&= \underbrace{ \int {q({\mathbf{x}} |{\mathbf{Y}})\log p({\mathbf{Y}}|{\mathbf{x}},{\mathbf{w}})d{\mathbf{x}} } }_{\text{Accuracy}} \nonumber\\
&\phantom{====} - \underbrace{\int {q({\mathbf{x}} |{\mathbf{Y}})\log \frac{{q({\mathbf{x}} |{\mathbf{Y}})}}{{p({\mathbf{x}} |{\mathbf{w}})}}d{\mathbf{x}} } }_{\text{Complexity}}
\label{eq:log_likelihood_accu_comp1}
\end{align}
Here, the first term on the right side of the last equation is an expectation of the log observation model by the recognition model, which represents the average goodness-of-fit of the observation model. This is sometimes referred to as the accuracy of the model. The second term is the KL divergence of the approximate posterior distribution and prior distribution, which is sometimes referred to as the complexity of the model. We discuss these two terms below.

First, let us consider the accuracy (the first term). In M-step, the parameters of the observation model are optimized by maximizing the accuracy. For example, parameters such as those of the basis functions are optimized, thereby improving the explanatory accuracy of the data. Not only that, Friston et al. hypothesized that actions are also chosen as to maximize the accuracy \cite{friston2006free,friston2010action,friston2012active,friston2012perceptions}. The inference problem in which action is involved is studied as an active inference problem, and it often refers to the selection of action according to maximizing the lower bound (or minimizing the variational free energy). In this case, the behavior of an agent is generated according to its generative model \cite{friston2009reinforcement,friston2012active,schwartenbeck2013exploration}. 
Actions selected by this principle are expected to alter the outside world so that it is more predictable for the agent. Note that when we introduce actions as a way of changing the data generation mechanism, the data distribution $\bar p(x)$ -- which has so far remained fixed -- is forced to change. However, organisms or agents are not able to access the true data distribution. Therefore, the only way to respond is by updating the generative model, and it is necessary for the generative model to include statements of how the generation of data is affected by actions. 



Now let’s consider the complexity (the second term). Complexity is minimized by bringing the prior distribution closer to the posterior distribution. In particular, when the approximate posterior distribution is an exact posterior distribution, we have the relation
\begin{align}
&\int {p({\mathbf{x}} |{\mathbf{Y}},{\mathbf{w}}) \log \frac{{p({\mathbf{x}} |{\mathbf{Y}},{\mathbf{w}})}}{{p({\mathbf{x}} | {\mathbf{w}})}}d{\mathbf{x}} } \nonumber\\
&\phantom{=}= {\rm{KL}}[p({\mathbf{x}} |{\mathbf{Y}},{\mathbf{w}}) ||p({\mathbf{x}} |{\mathbf{w}})],
\end{align}
which is called a Bayesian surprise \cite{itti2009bayesian}. This value quantifies how much the recognition model changes when the data ${\mathbf{Y}}$ is provided, given the parameter ${\mathbf{w}}$. The EM algorithm therefore performs approximate minimization of the Bayesian surprise. This means that surprises in the data are eliminated by adapting a prior distribution as a result of learning.

This process can represent the adaptability of organisms whereby the brain adapts to environmental factors or contexts, and incorporates them into its internal model. In particular, it was shown that the distance between the spontaneous firing activity and stimulus response activity of the visual cortex of ferrets decreases during the first 5 months of life \cite{berkes2011spontaneous}. The sampling hypothesis suggests that this observation manifests that the prior distribution and posterior distribution become closer due to learning by identifying spontaneous activity as samples from the prior distribution and stimulus-evoked activity as samples from the posterior distribution. The next section explains the adaptation process more generally from the viewpoint of information theory.

\section{Information-theoretic approaches to adaptation}\label{sec:information_theory}
In this section, we treat the learning process as communication through an information channel, and look at the relationship between the maximization of marginal likelihood in the previous section and maximization of information-theoretic quantities. Based on this, we will examine the relationship between the classical theory of perception \cite{barlow1961possible,linsker1988self,bell1995information} that is based on the information maximization principle (Infomax principle) and the approach with a generative model in the previous section. This can deepen the understanding of the relationship between the hypothesis of the brain as a nonlinear computing machine and that of the brain as an inference machine. See also \cite{kawato1993forward,dayan1995helmholtz,dayan2001theoretical,bogacz2017tutorial} for the relationship between the generative model and neural networks, and \cite{friston2010free,friston2012free,isomura2018measure} for the relationship between the free energy principle and information theory. 

\subsection{Maximization of mutual information and optimization of the generative model}
In this section, data ${\mathbf{Y}}$ is regarded as the input, and the parameter ${\mathbf{W}}$ of the generative model is regarded as the output. Here, we want to maximize the amount of mutual information between the random variable ${\mathbf{Y}}$ (data) and the model parameter ${\mathbf{W}}$. The mutual information can be expressed in terms of entropy and conditional entropy as follows:
\begin{align}
I({\mathbf{Y}};{\mathbf{W}}) &= H({\mathbf{Y}}) - H({\mathbf{Y}}|{\mathbf{W}}) \nonumber \\
&= H({\mathbf{Y}}) + E_{{\mathbf{Y}}} E_{{\mathbf{W}}|{\mathbf{Y}}} \log p({\mathbf{Y}}|{\mathbf{W}}).
\label{eq:mi1}
\end{align}
The first term on the right side in Eq.~\ref{eq:mi1} is the entropy of the data distribution $H({\mathbf{Y}}) \equiv - E_{{\mathbf{Y}}} \log \bar p({\mathbf{Y}}) $, and is constant when the data distribution is fixed. In this section, we will not consider changes to the data distribution caused by actions. Let us consider the second term on the right side. In the previous section, we considered the reverse process from the output ${\mathbf{W}}$ to the input ${\mathbf{Y}}$ as the data generation process in the internal model. Using latent variables, this path was obtained as the marginal distribution:
\begin{equation}
p({\mathbf{y}}|{\mathbf{w}}) = \int p({\mathbf{y}}|{\mathbf{x}},{\mathbf{w}}) p({\mathbf{x}}|{\mathbf{w}}) d{\mathbf{x}}.
\label{eq:communication_w_x}
\end{equation}
The second term in Eq.~\ref{eq:mi1} is the expectation of log of this function with respect to the joint distribution, $ p({\mathbf{y}},{\mathbf{w}})=\bar p({\mathbf{y}})p({\mathbf{w}}|{\mathbf{y}})$. 
The hierarchical model of the previous section possesses the parameter ${\mathbf{w}}$, which was learned from the data. This is a point estimate, which means that the distribution of ${\mathbf{w}}$ for a given set of data is given by $p({{\mathbf{w}}|{\mathbf{Y}}}) = \delta({{\mathbf{w}}-{\mathbf{W}}^{\star}})$, where $\delta(\cdot)$ is the Dirac delta function. Note that ${\mathbf{W}}^{\star}$ is a function of ${\mathbf{Y}}$.
In this way, the second term on the right side of Eq.~\ref{eq:mi1} becomes $E_{{\mathbf{Y}}} \log p({\mathbf{Y}}|{\mathbf{W}}^{\star})$. Further, by replacing the expected value of ${\mathbf{Y}}$ with the empirical distribution, we can use the data ${\mathbf{Y}}$ to obtain an estimate of the second term, which is the log marginal likelihood function. This means that we can expect the mutual information between ${\mathbf{Y}}$ and ${\mathbf{W}}$ to be maximized by adopting the maximum likelihood estimate ${\mathbf{W}}^{\ast}$ as a point estimation. 

\begin{figure}[t]
\begin{center}
\includegraphics[width=.5\textwidth]{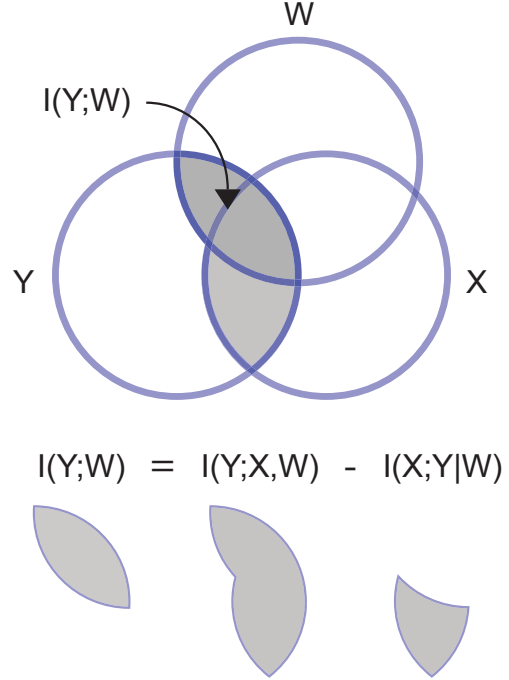}
\end{center}
\caption{Venn diagram and decomposition of mutual information}
\label{fig:venn_diagram}
\end{figure}

We note that, using a hidden state ${\mathbf{X}}$, the mutual information between ${\mathbf{Y}}$ and ${\mathbf{W}}$ is generally decomposed as
\begin{equation}
I({\mathbf{Y}};{\mathbf{W}}) = I({\mathbf{Y}};{\mathbf{X}}, {\mathbf{W}}) - I({\mathbf{X}};{\mathbf{Y}}|{\mathbf{W}}).
\label{eq:mi3}
\end{equation}
See the Venn diagram (Fig.~\ref{fig:venn_diagram}) for the visualization of this decomposition. From this decomposition, it is found that the mutual information $I({\mathbf{Y}};{\mathbf{W}})$ in the sensory data ${\mathbf{Y}}$ and the parameter ${\mathbf{W}}$ can be decomposed into two factors: the mutual information representing the encoding of sensory data ${\mathbf{Y}}$ by latent variables ${\mathbf{X}}$ and parameters ${\mathbf{W}}$ (first term), and the conditional mutual information between the latent variables ${\mathbf{X}}$ and data ${\mathbf{Y}}$, given prerequisite knowledge acquired by the parameter ${\mathbf{W}}$ (second term).\footnote{In this case, only the cost of encoding is considered. Tishby et al.’s information bottleneck theory, which builds on Shannon’s rate distortion theory, plays a complementary role in the theory of information maximization under the constraint that the cost of decoding is given for a particular generation model.} 
Below, we revisit the inference algorithm of the previous section, and clarify that they aim at optimizing these information-theoretic quantities. 

Based on Eq.~\ref{eq:log_likelihood_accu_comp1}, the marginal likelihood can be expressed as
\begin{align}
\log p({\mathbf{Y}}|{\mathbf{W}}^{\star})
&= E_{{\mathbf{X}}|\mathbf{Y},{\mathbf{W}}^{\star}} \log p({\mathbf{Y}}|{\mathbf{X}}, {\mathbf{W}}^{\star}) \nonumber\\
&\phantom{=} - {\rm{KL}} [p({\mathbf{x}}|{\mathbf{Y}},{\mathbf{W}}^{\star}) \| p({\mathbf{x}}|{\mathbf{W}}^{\star})],
\label{eq:log_likelihood_accu_comp}
\end{align}
where it is assumed that an exact posterior distribution has been obtained. The first term represents the accuracy, and the second term represents the Bayesian surprise. We will confirm that maximization of the accuracy and minimization of the Bayesian surprise correspond to maximization of the first term and minimization of the second term in Eq.~\ref{eq:mi3}, respectively. 

First we note that $E_{{\mathbf{W}}|{\mathbf{Y}}}$ represents expectation by the posterior distribution of ${\mathbf{W}}$, namely $p({\mathbf{w}}|{\mathbf{Y}})=\delta({\mathbf{w}}-{\mathbf{W}}^{\star})$. Therefore, the log marginal likelihood can be expressed as $\log p({\mathbf{Y}}|{\mathbf{W}}^{\star})=E_{{\mathbf{W}}|{\mathbf{Y}}} \log p({\mathbf{Y}}|{\mathbf{W}})$. 
By taking the expectation by $E_{{\mathbf{Y}}}$, we can see that the left side of Eq.~\ref{eq:log_likelihood_accu_comp} corresponds to
\begin{equation}
E_{{\mathbf{Y}}} E_{{\mathbf{W}}|{\mathbf{Y}}} \log p({\mathbf{Y}}|{\mathbf{W}}) = - H({\mathbf{Y}}|{\mathbf{W}}).
\end{equation}
Similarly, expectation of the first term on the right side by data distribution is
\begin{equation}
E_{{\mathbf{Y}}} E_{{\mathbf{W}}|{\mathbf{Y}}} E_{{\mathbf{X}}|{\mathbf{Y}},{\mathbf{W}}} \log p({\mathbf{Y}}|{\mathbf{X}}, {\mathbf{W}} ) = - H({\mathbf{Y}}|{\mathbf{X}}, {\mathbf{W}} ).
\end{equation}
Second, the expectation of the Bayesian surprise (the second term on the right side) by the data distribution becomes the conditional mutual information:
\begin{equation}
E_{{\mathbf{Y}}} E_{{\mathbf{W}}|{\mathbf{Y}}} {\rm{KL}} [p({\mathbf{x}}|{\mathbf{Y}},{\mathbf{W}}) \| p({\mathbf{x}}|{\mathbf{W}})] = I({\mathbf{X}}; {\mathbf{Y}}|{\mathbf{W}}). 
\label{eq:mi_cond}
\end{equation}
Note that the mutual information $X$ and $Y$ available under $Z$ is computed from the KL divergence:\footnote{In general, the mutual information of continuous random variables X and Y is given by  
\begin{align}
I(X;Y) &= \iint p(x,y) \log \frac{p(x,y) }{p(x)p(y)} dx dy\nonumber\\
&= \iint p(x) p(y|x) \log \frac{p(y|x) }{p(y)} dx dy \nonumber\\
&=E_{X} {\rm{KL}}[p(y|X) || p(y) ]
\end{align}
}
\begin{align}
&I(X;Y|Z) \nonumber\\
&\phantom{=}= \iiint p(x,y,z) \log \frac{p(x,y|z) }{p(x|z)p(y|z)} dx dy dz\nonumber\\
&\phantom{=}=E_{X} E_{Z|X} {\rm{KL}}[p(y|X,Z) || p(y|Z) ]
\end{align}
The mutual information for discrete random variables can also be defined in the same way by replacing the integral with a summation. Eq.~\ref{eq:mi_cond} is an extension of this formula to multivariate variables. In other words, Bayesian surprise is an estimate of conditional mutual information. Taken together, expectation of Eq.~\ref{eq:log_likelihood_accu_comp} by $E_{{\mathbf{Y}}}$ is simplified to
\begin{equation}
-H({\mathbf{Y}}|{\mathbf{W}}) = -H({\mathbf{Y}}|{\mathbf{X}}, {\mathbf{W}}) - I({{\mathbf{X}}; \mathbf{Y}}|{\mathbf{W}}).
\label{eq:mi5}
\end{equation}
By adding the entropy of the data $H({\mathbf{Y}})$ on both sides of Eq.~\ref{eq:mi5}, we obtain Eq.~\ref{eq:mi3}.\footnote{
Given the hierarchical model of the brain (Eq.~\ref{eq:generative_model_of_brain}), Eq.~\ref{eq:mi3} can be further written as  
\begin{equation}
I({\mathbf{Y}};{\boldsymbol{\phi}},{\boldsymbol{\lambda}}) = I({\mathbf{Y}};{\mathbf{X}}, {\boldsymbol{\phi}}) - I({{\mathbf{X}}; \mathbf{Y}}, {\boldsymbol{\phi}}|{\boldsymbol{\lambda}}).
\label{eq:mi6}
\end{equation}
This is obtained as follows. Under the assumption of the hierarchical model shown in Fig.~\ref{fig:graphical_model}, we obtain $I({\mathbf{Y}};{\mathbf{X}}, {\boldsymbol{\phi}},{\boldsymbol{\lambda}})=I({\mathbf{Y}};{\mathbf{X}}, {\boldsymbol{\phi}})$ since ${\mathbf{X}}$ is conditionally independent of ${\boldsymbol{\lambda}}$. Using the general decomposition rule of Eq.~\ref{eq:mi3} on the distributions conditional on ${\boldsymbol{\lambda}}$, we have $I({\mathbf{Y}};{\mathbf{X}} | {\boldsymbol{\phi}},{\boldsymbol{\lambda}})=I({\mathbf{X}};{\mathbf{Y}},{\boldsymbol{\phi}}|{\boldsymbol{\lambda}}) - I({\mathbf{X}}; {\boldsymbol{\phi}}|{\boldsymbol{\lambda}})=I({\mathbf{X}};{\mathbf{Y}},{\boldsymbol{\phi}}|{\boldsymbol{\lambda}})$. Here we used $I({\mathbf{X}}; {\boldsymbol{\phi}}|{\boldsymbol{\lambda}})=0$ because ${\mathbf{X}}$ and ${\boldsymbol{\phi}}$ are independent if ${\mathbf{Y}}$ is marginalized.
}

By comparing Eq.~\ref{eq:log_likelihood_accu_comp} with Eq.~\ref{eq:mi3}, we can clearly understand objectives that optimization of each term in Eq.~\ref{eq:log_likelihood_accu_comp} aims at during the learning. First, while keeping in mind the existence of an unknown constant $H({\mathbf{Y}})$, maximizing the first term on the right side of Eq.~\ref{eq:log_likelihood_accu_comp} (accuracy) maximizes the (estimated) mutual information related to the encoding from ${\mathbf{Y}}$ to ${\mathbf{X}}$ and ${\mathbf{W}}$. 
Second, minimization of complexity/Bayesian surprise in the second term of Eq.~\ref{eq:log_likelihood_accu_comp} minimizes the (estimated) mutual information between the neural activity ${\mathbf{X}}$ and the data ${\mathbf{Y}}$, conditional on the parameter ${\mathbf{W}}$ of the generative model. This means that the information of the input data ${\mathbf{Y}}$ is absorbed during the process of learning the parameter ${\mathbf{W}}$, and as a result, the neural activity ${\mathbf{X}}$ no longer has any information about the input data ${\mathbf{Y}}$ other than what is held by ${\mathbf{W}}$.

\subsection{Infomax principle for optimization of nonlinear networks}
In this subsection, we review the classical theory of sensory perception. Based on this, we will consider its relation to the approach based on the generative model. Here we assume noiseless communication between the input $\mathbf{Y}$ and the output $\mathbf{X}$, and that they have the same dimension.
For the noiseless channel, we can represent ${\mathbf{X}}$ as a nonlinear function of ${\mathbf{Y}}$, using ${\mathbf{x}}={\mathbf{f}}({\mathbf{y}}; {\boldsymbol{\varphi}})$, for which a neural network can be used. Here ${\boldsymbol{\varphi}}$ are the parameters that define the nonlinear function (e.g., weights of a neural network), and comprise receptive fields of the neurons. 

The adaptation of the neural activity ${\mathbf{X}}$ to the external input ${\mathbf{Y}}$ by maximizing the mutual information $I({\mathbf{Y}};{\mathbf{X}})$ is known as the infomax principle \cite{linsker1988self}. This mutual information is expressed in terms of entropy as
\begin{equation}
I({\mathbf{Y}};{\mathbf{X}}) = H({\mathbf{X}}) - H({\mathbf{X}}|{\mathbf{Y}}).
\label{eq:mi2}
\end{equation}
In a noiseless channel, the second term of Eq.~\ref{eq:mi2} is not dependent on the nonlinear function ${\mathbf{f}}({\mathbf{y}}; {\boldsymbol{\varphi}})$, so it can be ignored in optimization.\footnote{The conditional entropy is 0 for discrete distributions, or negative infinity for continuous distributions if the channel is noiseless.} Therefore, maximizing the mutual information is equivalent to maximizing the entropy of the output ${\mathbf{X}}$. 

The method of optimizing parameter ${\boldsymbol{\varphi}}$ based on this principle is known as independent component analysis (ICA) used for blind signal separation\cite{bell1995information}. That is, the parameter ${\boldsymbol{\varphi}}$ is updated using the following gradient,
\begin{equation}
\frac{\partial I({\mathbf{Y}};{\mathbf{X}}) } {\partial {\boldsymbol{\varphi}}} = \frac{\partial H({\mathbf{X}}) } {\partial {\boldsymbol{\varphi}}}.
\label{eq:der_mi_Y_X}
\end{equation}
Here, the entropy of ${\mathbf{X}}$ due to the stochastic data ${\mathbf{Y}}$ is given by
\begin{equation}
H({\mathbf{X}}) = H({\mathbf{Y}}) + E_{{\mathbf{Y}}} \log \det \left. { \frac{\partial {\mathbf{f}}({\mathbf{y}}; {\boldsymbol{\varphi}})}{\partial {{\mathbf{y}}}} } \right|_ {{\mathbf{y}}={\mathbf{Y}} },
\label{eq:entropy_X}
\end{equation}
because $p({\mathbf{y}}) = p({\mathbf{x}}) \left| \frac{\partial {\mathbf{x}}}{\partial {\mathbf{y}}} \right|$, where $\frac{\partial {\mathbf{x}}}{\partial {\mathbf{y}}} =  \frac{\partial {\mathbf{f}}({\mathbf{y}}; {\boldsymbol{\varphi}})}{\partial {\mathbf{y}}}$ is a Jacobi matrix used for the change of variables, and $ \left| \cdot  \right| = \det \cdot$ denotes a determinant. Using Eqs.~\ref{eq:der_mi_Y_X} and \ref{eq:entropy_X}, and replacing the expectation of data with a sample ${\mathbf{Y}}$, the learning rule is obtained as
\begin{equation}
\frac{d {\boldsymbol{\varphi}} } {d t} \propto \frac{\partial } {\partial {\boldsymbol{\varphi}}} \log \det \left. { \frac{\partial {\mathbf{f}}({\mathbf{y}}; {\boldsymbol{\varphi}})}{\partial {{\mathbf{y}}}} } \right|_ {{\mathbf{y}}={\mathbf{Y}} }.
\label{eq:learning_infomax}
\end{equation}

The mutual information maximization based on maximizing the output entropy follows the framework of the efficient coding hypothesis proposed by Horace Barlow \cite{barlow1961possible,barlow1972single}. On the assumption of a noiseless channel, Barlow proposed the principle of maximizing entropy of ${\mathbf{X}}$ (i.e., activity of neurons) as a goal of encoding sensory data such as tactile and vision. The efficient coding hypothesis is also called the redundancy reduction hypothesis since it involves eliminating redundancy from data as shown below. 

To explain this, let's see that the multivariate entropy can be decomposed as follows.
\begin{equation}
H({\mathbf{X}}) = - I({X}_1;{X}_2;\ldots;{X}_d) + \sum_{i=1}^{d}H({X}_i).
\end{equation}
Here, the first term on the right side is the multivariate mutual information, defined as $I({X}_1;\ldots;{X}_d)= {\rm KL} [p({\mathbf{X}} || \Pi_{i=1}^{d} p({X}_i )]$. Two conclusions can be derived from this equation. From the first term, we can see that in order to increase the mutual information at the inputs and outputs, it is better to have a smaller quantity of mutual information between the outputs. That is, the parameter ${\boldsymbol{\varphi}}$ should be adjusted so that the output variables become independent random variables. This sort of encoding, which aggregates external signals and converts them into independent representations, is sometimes called factorial coding.

Next, from the second term, it can be seen that in order to increase the mutual information between the input and output, the entropy of each random variable of the output should be increased. Since the entropy is maximized when the random variables have a uniform distribution, the parameter ${\boldsymbol{\varphi}}$ should be adjusted so as to obtain a uniform output from the nonlinear function ${\mathbf{X}}={\mathbf{f}}({\mathbf{Y}}; {\boldsymbol{\varphi}})$. For example, it is the most appropriate if the activity of individual neurons is covered uniformly over its dynamic range. Consider a one-dimensional sensory input $\bar p({{y}})$ for simplicity. If we wish to construct a uniform distribution that has an upper limit as an output ${{X}}={{f}}({{Y}}; {{\varphi}})$, we should use the nonlinear function ${{f}}({{y}}; {{\varphi}}) = \int_{-\infty}^{y} \bar p({y'}) d{y'}$ or a constant multiple thereof. This is because substituting a random variable ${{Y}}$ whose distribution is $\bar p({{y}})$ to its own cumulative distribution function results in a uniform distribution of values over the range $[0,1]$.\footnote{If $X$ is a random variable following the probability distribution function $F_X(x)=\int_{-\infty}^{x}f_X(s) ds$, then $U=F_X(X)$ has a uniform distribution. It is also used as a way of creating arbitrary sample distributions from uniform random numbers (inverse function method): $X=F_X^{-1}(U)$} This nonlinear transformation is sometimes called histogram equalization.

\begin{figure}[t]
\begin{center}
\includegraphics[width=.5\textwidth]{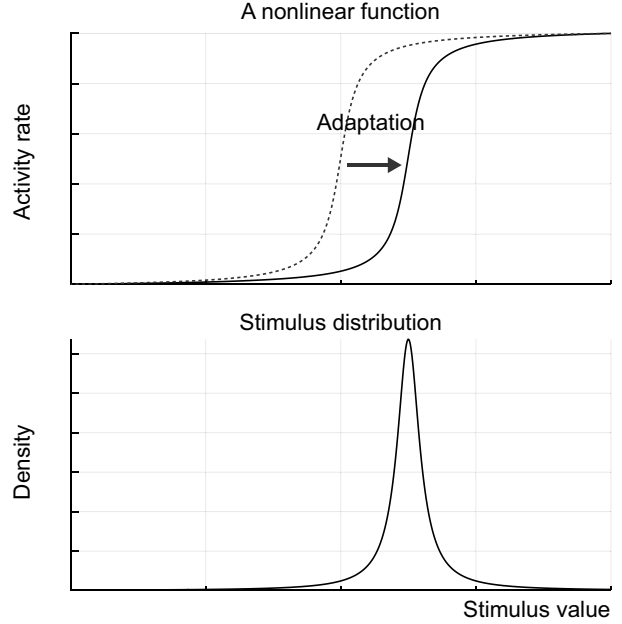}
\end{center}
\caption{Nonlinear functions and adaptation}
\label{fig:nonlinearfunc}
\end{figure}

According to the efficient coding hypothesis, organisms are thought to adapt its nonlinear activation functions of neurons to the distribution function of the input data so they can use the output independently and uniformly over its dynamic range. In particular, horizontal movements or multiplications are added to basic nonlinear input/output functions to rapidly adapt to the environment. This process is called gain control (see Fig.~\ref{fig:nonlinearfunc}). In insect retinal and olfactory nerve cells \cite{laughlin1981simple,olsen2010divisive} and mammalian visual neurons \cite{shapley1978effect,laughlin1989role,ohzawa1982contrast}, the gain control is performed to adapt to changes in the environment. Furthermore, the gain control explains higher-level cognitive functions of a brain such as attention and coordinate transformation of viewpoint \cite{reynolds2000attention,salinas2001gain,reynolds2009normalization,carandini2012normalization}. 

While the gain control allows fast adaptation to changing environment, nonlinear functions themselves must be acquired from the data during learning processes in a longer time-scale. To capture the non-Gaussian nature of sensory inputs, it is necessary to adaptively generate a nonlinear function that matches up the higher order statistics of the distribution. For the distribution whose dimension is higher than two, the higher-order correlations (higher-order statistics among the variables) in the distribution must be captured. In this way, it is hypothesized that nonlinear functions that are expressed through the nonlinearity of neurons (nonlinear summation of synaptic inputs at dendrites, firing characteristics based on threshold mechanisms, and network dynamics) are adapted to the characteristics of the distribution of input data \cite{nadal1994nonlinear,schwartz2001natural}. 

Finally, we examine how the above arguments on the infomax principle can be generalized to the approach with the generative model. We investigated optimization of the nonlinear function ${\mathbf{x}}={\mathbf{f}}({\mathbf{y}}; {\boldsymbol{\varphi}})$, assuming a noiseless channel and equal dimensions for ${\mathbf{y}}$ and ${\mathbf{x}}$. For simplicity, here we assume that the nonlinear function is bijective functions (one-to-one functions), and define the inverse function as ${\mathbf{y}}={\mathbf{g}}({\mathbf{x}}; {\boldsymbol{\phi}})$. The parameter ${\boldsymbol{\phi}}$ represents how the data is constructed from the neural activity, and is called the basis function or projection field \cite{dayan2001theoretical}. This is in contrast to the parameter ${\boldsymbol{\varphi}}$ that comprises the receptive field of neurons, which explains neural activity in terms of the sensory inputs.\footnote{If the functions and parameters are linear (${\mathbf{Y}}={\mathbf{X}} {\boldsymbol{\varphi}}$ and ${\mathbf{X}}={\mathbf{Y}} {\boldsymbol{\phi}}$), we obtain the relation, ${\boldsymbol{\varphi}} {\boldsymbol{\phi}} = \mathbf{I}$.} 

For the noiseless channel, the observation model is written as $p({\mathbf{Y}}|{\mathbf{x}},{\boldsymbol{\phi}})=\delta({\mathbf{Y}} - {\mathbf{g}}({\mathbf{x}}; {\boldsymbol{\phi}}))$. Then the marginal likelihood function is written as\footnote{Here, with the change of a variable $\mathbf{y}={\mathbf{g}}({\mathbf{x}}; {\boldsymbol{\phi}})$, we have 
\begin{align}
p({\mathbf{Y}}|{\mathbf{w}}) 
&= \int \delta({\mathbf{Y}}-{\mathbf{g}}({\mathbf{x}};{\boldsymbol{\phi}})) \pi({\mathbf{x}}|{\boldsymbol{\lambda}}) d{\mathbf{x}} \nonumber\\
&= \int \delta({\mathbf{Y}}-\mathbf{y}) \pi({\mathbf{g}}^{-1}({\mathbf{y}}; {\boldsymbol{\phi}})|{\boldsymbol{\lambda}}) \left| \frac{\partial {\mathbf{x}}}{\partial {\mathbf{y}}} \right| d{\mathbf{y}} \nonumber\\
&= \pi({\mathbf{g}}^{-1}({\mathbf{Y}}; {\boldsymbol{\phi}})|{\boldsymbol{\lambda}}) \left. \det \frac{\partial {\mathbf{x}}}{\partial {\mathbf{y}}} \right|_{\mathbf{y}=\mathbf{Y}} \nonumber\\
&= \pi({\mathbf{f}}({\mathbf{Y}};{\boldsymbol{\varphi}}) |{\boldsymbol{\lambda}}) \cdot \det \left.\frac{{\partial \mathbf{f}}({\mathbf{y}};{\boldsymbol{\varphi}})}{\partial {\mathbf{y}}}\right|_{{\mathbf{y}}={\mathbf{Y}}}.
\end{align}
}
\begin{align}
p({\mathbf{Y}}|{\mathbf{w}}) 
&= \int \delta({\mathbf{Y}}-{\mathbf{g}}({\mathbf{x}};{\boldsymbol{\phi}})) \pi({\mathbf{x}}|{\boldsymbol{\lambda}}) d{\mathbf{x}} \nonumber\\
&= \pi({\mathbf{X}})  \cdot  \det \left.\frac{{\partial \mathbf{f}}({\mathbf{y}};{\boldsymbol{\varphi}})}{\partial {\mathbf{y}}}\right|_{{\mathbf{y}}={\mathbf{Y}}}.
\label{eq:marginal_likelhood_noiseless}
\end{align}
Here the prior distribution for ${\mathbf{X}}$ is denoted by $\pi(\cdot)$ to avoid confusion. In the right hand side of this equation, the sensory data ${\mathbf{Y}}$ is projected to the neural activity ${\mathbf{X}}$ by the nonlinear function ${\mathbf{x}}={\mathbf{f}}({\mathbf{y}};{\boldsymbol{\varphi}})$, and the likelihood of the sensory data ${\mathbf{Y}}$ is now evaluated using the neural activity ${\mathbf{X}}$ with respect to its prior distribution. Under the maximum likelihood principle, learning of the nonlinear function is performed so that the neural activity ${\mathbf{X}}$ fits to the prior distribution. At the same time, the parameter ${\boldsymbol{\lambda}}$ of the prior distribution is optimized to fit the neural activity. These learning processes correspond to optimization of the observation model and the prior distribution discussed in the previous section, respectively. Note that if we assume a flat prior, we obtain the same gradient in Eq.~\ref{eq:learning_infomax} for the infomax principle from the principle of maximizing the log marginal likelihood. This indicates that, in the approach with the generative model, the optimal nonlinear function ${\mathbf{x}}={\mathbf{f}}({\mathbf{y}};{\boldsymbol{\varphi}})$ is modulated by the prior distribution, which can be interpreted as the gain modulation.

One can arrive at the generative model investigated in the previous section by generalizing the above deterministic observation model to noisy observation models togather with non-bijective nonlinear functions. Often the mean $\boldsymbol{\mu}$ of the observation model is modeled as $\boldsymbol{\mu}= \mathbf{g}({\mathbf{X}} {\boldsymbol{\phi}})$, where ${\boldsymbol{\phi}}$ plays the role of linear basis functions. Such a nonlinear function $\mathbf{g}(\cdot)$ is called a link function in statistics. 
Olshausen \& Filed introduced a sparse prior distribution to the latent variables of the Gaussian observation model with a linear link function, and make the basis functions of the observation model learned from natural images. In this model, the dimension of latent variables ${\mathbf{Y}}$ and corresponding basis functions are much larger than the number of pixels in ${\mathbf{Y}}$ (overcomplete model). They then found that the model learns high-dimensional correlation structures such as lines and edges that often appear in natural images \cite{olshausen1996emergence,olshausen1997sparse}.

An alternative approach to augment the noiseless model to a noisy model is to directly construct the approximate posterior distribution $q({\mathbf{x}}|{\mathbf{Y}})$ by adding noise to the nonlinear function ${\mathbf{x}}={\mathbf{f}}({\mathbf{y}}; {\boldsymbol{\varphi}})$. In the original variational auto-encoder model, not only the mean but variance of the approximated Gaussian posterior distribution is learned by neural networks \cite{kingma2017variational}. Contrary to the overcomplete representation, the dimension of the latent variable ${\mathbf{X}}$ is smaller than that of the observation ${\mathbf{Y}}$; therefore the posterior called encoder maps the data into latent variables in a smaller subspace whereas the observation model called decoder maps the latent variables to the data, which has a larger dimension. A scheme different from the EM algorithm was developed to learn the posterior and the generative model simultaneously using the neural networks. 
Common to all approaches, it is known that neither of these approaches can fully eliminate redundancy in simple nonlinear functions or non-Gaussian distributions \cite{schwartz2001natural,simoncelli2001natural,maboudi2016representation}, so efforts are being made to grasp the input correlation structure with deeply hierarchical models.

\section{Thermodynamics of adaptation}\label{sec:thermodynamics}
In Section \ref{sec:approximate_inference}, we introduced the EM algorithm as a learning method for hierarchical models based on Helmholtz’s epistemology. This approach can be extended to variational inference using variational approximation methods. The variational inference is a technique derived from mean field approximation in statistical physics that provides approximate solutions to physical models of materials involving complex interactions among their elements (e.g., Ising model). Therefore its mathematical framework is closely related to statistical physics and thermodynamics.\footnote{As a physiologist, Helmholtz himself developed a theory of visual perception (the Young–Helmholtz theory of trichromatic color vision), and as a physicist, he contributed to establishing the first law of thermodynamics and the theory of free energy in chemical reactions.} However, it is not always clear how the laws of thermodynamics relate to recognition and learning. 

In this section, we will look at the learning process from a thermodynamic viewpoint based on the principle of entropy maximization, and we will discuss dynamics of learning by defining energy and free energy of a recognition model using an exponential family distribution. Based on this, we explain the relation between causal statistical learning and the second law of thermodynamics. A similar treatment to the neural dynamics presented in this section can be found in the thermodynamic analysis of neural population activity in Shimazaki 2015, 2018 \cite{shimazaki2015neurons,shimazaki2018neural} to which the reader is referred. Methods for calculating thermodynamic quantities of neural populations from actual recordings of time-series neural spike data, together with the results of these methods, can be found in Tkačik et al. 2015 \cite{tkavcik2015thermodynamics}, Donner et al. 2017 and Gaudreault et al. 2018 \cite{donner2017approximate,gaudreault2018state}.

\subsection{Maximum entropy model}
Before entering the discussion, we start with a description of the terminology. Based on Eq.~\ref{eq:lowerbound_decomposition}, the lower bound $\mathcal{L}[q,p]$ with a recognition model $q({\mathbf{x}}|{\mathbf{Y}})$ can be expressed as follows: 
\begin{equation}
\mathcal{L}[q,p] = S - \int {q({\mathbf{x}}|{\mathbf{Y}}) [- \log p({\mathbf{Y}},{\mathbf{x}} |{\mathbf{w}}) ] d{\mathbf{x}} }.
\label{eq:elbo_entropy_qfunc2}
\end{equation}
In this section, the Shannon entropy of the recognition model is represented by $S$ ($S \equiv H[q({\mathbf{x}} |{\mathbf{Y}})] $). The second term is the negative of the $\mathcal{Q}$-function ($-\mathcal{Q}$), which is as follows:
\begin{equation}
\int {q({\mathbf{x}}|{\mathbf{Y}}) [- \log p({\mathbf{Y}},{\mathbf{x}} |{\mathbf{w}}) ] d{\mathbf{x}} } = \mathcal{E}.
\label{eq:constraint}
\end{equation}
As shown later, $\mathcal{E}$ is a quantity that can be referred to as the energy of the recognition model. These terms are used to bound the negative marginal likelihood function as
\begin{equation}
- \log p({\mathbf{Y}}|{\mathbf{w}}) \leq \mathcal{E}-S \equiv F,
\end{equation}
where, in statistical physics, the left side is called free energy and the right side $F = \mathcal{E} - S$ is called the variational free energy. The variational free energy is the negative lower bound ($F = - \mathcal{L}[q,p]$), and the problem of maximizing the lower bound that was discussed in the section of approximate inference can be substituted with the problem of minimizing the variational free energy. In the following, a thermodynamic approach to a recognition model will be introduced to better understand dynamics of the entropy, energy, and free energy (not variational free energy) for a recognition model, induced by learning.

For this goal, we will construct a recognition model from the maximum entropy principle. Consider the problem of maximizing the entropy of a recognition model $q({\mathbf{x}}|{\mathbf{Y}})$ given an expected value of a generative model. This is an entropy maximization problem with constraints, which can be solved by the method of Lagrange multipliers. In this method, the constrained maximization problem is replaced with maximization of a new function that includes the constraints. More specifically, we maximize the following Lagrange function (Lagrangian):
\begin{align}
&\mathcal{\tilde L}_{\beta}[q]
=- \int q({\mathbf{x}}|{\mathbf{Y}}) \log q({\mathbf{x}}|{\mathbf{Y}}) d{\mathbf{x}} \nonumber\\
&\phantom{=} - \beta \left\{- \int {q({\mathbf{x}}|{\mathbf{Y}}) \log p({\mathbf{Y}},{\mathbf{x}} |{\mathbf{w}}) d{\mathbf{x}} } - \mathcal{E} \right\} \nonumber\\
&\phantom{==} + a \left\{ \int {q({\mathbf{x}}|{\mathbf{Y}})} d{\mathbf{x}} - 1\right\}.
\label{eq:lagrangian1}
\end{align}
where $\beta,$ and $a$ are Lagrange multipliers. The last term is a constraint due to the fact that the recognition model is a density or probability function. According to the variational principle, we need to obtain the distribution by maximizing Eq.~\ref{eq:lagrangian1}. The variation with respect to the distribution $q({\mathbf{x}}|{\mathbf{Y}})$ is given by
\begin{align}
\frac{\delta \mathcal{\tilde L}_{\beta}[q] } {\delta q}
&=\int \delta q \, [ -1 - \log q({\mathbf{x}}|{\mathbf{Y}}) \nonumber\\
&\phantom{=} + \beta { \log p({\mathbf{Y}},{\mathbf{x}} |{\mathbf{w}}) } + a ] d{\mathbf{x}}.
\end{align}
Thus it is found that the distribution obeys the following exponential family distribution:
\begin{equation}
q({\mathbf{x}}|{\mathbf{Y}}) = \frac{1}{Z_{\beta}({\mathbf{Y}})} e^{-\beta \left\{ - \log p({\mathbf{Y}},{\mathbf{x}} |{\mathbf{w}}) \right\}}.
\label{eq:maxent}
\end{equation}
The term $Z_{\beta}({\mathbf{Y}})(=e^{a-1})$ is called a normalization term or partition function, and is given by 
\begin{equation}
Z_{\beta}({\mathbf{Y}}) = \int e^{-\beta \left\{ - \log p({\mathbf{Y}},{\mathbf{x}} |{\mathbf{w}}) \right\}} d{\mathbf{x}}.
\label{eq:posterior_maxent}
\end{equation}
The Lagrange multiplier $\beta$ is obtained at the maxima of the Lagrangian ($\frac{\partial \mathcal{\tilde L}_{\beta}[q] } {\partial \beta}=0$), which is given by Eq.~\ref{eq:constraint}. That is, $\beta$ is chosen to satisfy the constraints of Eq.~\ref{eq:constraint}. Note that in this section, we consider the set of parameters as $\mathbf{w}=\{{\boldsymbol{\varphi}}, {\boldsymbol{\omega}}\}$ by excluding $\boldsymbol{\beta}$, and separately introduce $\beta$ as a Lagrange multiplier that is a parameter of the posterior distribution. 

The maximum entropy method is a method that maximizes the entropy of a distribution while applying specific constraints. It can be used to obtain distributions by eliminating statistical structures other than constraints. As seen above, its distribution is expressed by an exponential family distribution from the definition of entropy. In statistical physics, this distribution is called the Gibbs or Boltzmann distribution, the exponent of the exponential function $\mathcal{H}({\mathbf{x}}) \equiv - \log p({\mathbf{Y}},{\mathbf{x}} |{\mathbf{w}}) $ is called the Hamiltonian, and the expected value of the Hamiltonian is called the energy as given by Eq.~\ref{eq:constraint}. Also, $\beta$ is called the inverse temperature, and $T\equiv1/\beta$ is called the temperature. When $\beta=1$, the distribution that maximizes the entropy according to Eq.~\ref{eq:maxent} becomes
\begin{equation}
q_{\beta=1}({\mathbf{x}}|{\mathbf{Y}}) = \frac{p({\mathbf{Y}}, {\mathbf{x}} |{\mathbf{w}})}{Z_{\beta=1}({\mathbf{Y}})} = p({\mathbf{x}} |{\mathbf{Y}},{\mathbf{w}}),
\label{eq:posterior_maxent_exact}
\end{equation}
which is an exact posterior distribution.

\subsection{Law of conservation of entropy for a recognition model}
The generative model can be divided into an observation model and a prior distribution. To manipulate the entropy of the recognition model more precisely than in the above formula, we now consider a recognition model, for which the constraints are stated in more details. More specifically, we search for the recognition model whose entropy is maximized under the following constraints:
\begin{align}
\langle -\log p({\mathbf{x}} |{\boldsymbol{\omega}}) \rangle &= U, \\
\langle -\log p({\mathbf{Y}}|{\mathbf{x}}, {\boldsymbol{\phi}}) \rangle &= V.
\end{align}
where $\langle \cdot \rangle$ is the expected value of the recognition model. In this case, the Lagrangian can be written as follows:
\begin{align}
&\mathcal{\tilde L}_{\beta,\alpha}[q]
=- \int q({\mathbf{x}}|{\mathbf{Y}}) \log q({\mathbf{x}}|{\mathbf{Y}}) d{\mathbf{x}} \nonumber\\
&\phantom{=} - \beta \left\{- \int {q({\mathbf{x}}|{\mathbf{Y}}) \log p({\mathbf{x}} |{\boldsymbol{\omega}})d{\mathbf{x}} } - U \right\} \nonumber\\
&\phantom{==} - \alpha \left\{- \int {q({\mathbf{x}}|{\mathbf{Y}}) \log p({\mathbf{Y}}|{\mathbf{x}},{\boldsymbol{\phi}})d{\mathbf{x}} } - V \right\} \nonumber\\
&\phantom{===} + a \left\{ \int {q({\mathbf{x}}|{\mathbf{Y}})} d{\mathbf{x}} - 1\right\},
\end{align}
where $\beta$, $\alpha$, and $a$ are Lagrange multipliers.\footnote{In the earlier equations, $\beta$ was a Lagrange multiplier for the generative model. However, it is now a Lagrange multiplier for the prior distribution. Therefore, $\beta$ is no longer a parameter that controls energy. If we introduce $\alpha = \beta f$, $\beta$ controls the energy.} By examining variations in the recognition model in the same way as before, we obtain
\begin{align}
&\frac{\delta \mathcal{\tilde L}_{\beta,\alpha}[q]} {\delta q}
=\int \delta q \, [ -1 - \log q({\mathbf{x}}|{\mathbf{Y}}) \nonumber\\
&\phantom{=} + \beta { \log p({\mathbf{x}} |{\boldsymbol{\omega}})} + \alpha {\log p({\mathbf{Y}}|{\mathbf{x}},{\boldsymbol{\phi}}) } + a ] d{\mathbf{x}}.
\end{align}
Consequently, the recognition model obeys the following exponential family distribution:
\begin{equation}
q({\mathbf{x}}|{\mathbf{Y}}) = \frac{1}{Z_{\beta,\alpha}({\mathbf{Y}})} e^{\beta \log p({\mathbf{x}} |{\boldsymbol{\omega}}) + \alpha \log p({\mathbf{Y}}|{\mathbf{x}},{\boldsymbol{\phi}})}.
\label{eq:posterior_maxent2}
\end{equation}
The partition function is given by
\begin{equation}
Z_{\beta,\alpha}({\mathbf{Y}}) = \int e^{\beta \log p({\mathbf{x}} |{\boldsymbol{\omega}}) + \alpha \log p({\mathbf{Y}}|{\mathbf{x}},{\boldsymbol{\phi}})} d{\mathbf{x}}.
\label{eq:partition_func2}
\end{equation}
The Lagrange multipliers $\beta$ and $\alpha$ are chosen to satisfy the constraints. When $\beta=1$ and $\alpha=1$, the recognition model becomes an exact posterior distribution.

The recognition model of Eq.~\ref{eq:posterior_maxent2} can be regarded as an exponential family distribution where $-\beta$ and $-\alpha$ are canonical parameters, and $- \log p({\mathbf{x}} |{\boldsymbol{\omega}})$ and $- \log p({\mathbf{Y}}|{\mathbf{x}},{\boldsymbol{\phi}})$ are features. We can therefore derive a number of important relations. First, the logarithm of the partition function forms a cumulant generating function. As a result, the first derivative of the log partition function by the canonical parameter gives the expected value of the feature by the recognition model. That is,
\begin{align}
\frac{{\partial \log Z_{\beta,\alpha}({\mathbf{Y}})}}{{\partial (-\beta)}} &= \langle -\log p({\mathbf{x}} |{\boldsymbol{\omega}}) \rangle \nonumber\\
\frac{{\partial \log Z_{\beta,\alpha}({\mathbf{Y}})}}{{\partial (-\alpha)}} &= \langle -\log p({\mathbf{Y}}|{\mathbf{x}}, {\boldsymbol{\phi}}) \rangle.
\end{align}
Here, by defining a function
\begin{equation}
\mathcal{G}(\beta,\alpha) \equiv - \log Z_{\beta,\alpha}({\mathbf{Y}}),
\end{equation}
they can be simplified to
\begin{equation}
\frac{{\partial \mathcal{G}(\beta,\alpha)}}{{\partial \beta}} = U, \ \ \ \frac{{\partial \mathcal{G}(\beta,\alpha)}}{{\partial \alpha}} = V.
\label{eq:partition_expectation}
\end{equation}
Next, the entropy of the recognition model can be calculated as
\begin{align}
S(U,V) &= \langle -\log q({\mathbf{x}}|{\mathbf{Y}}) \rangle \nonumber\\
&= \beta \langle -\log p({\mathbf{x}} |{\boldsymbol{\omega}}) \rangle + \alpha \langle -\log p({\mathbf{Y}}|{\mathbf{x}},{\boldsymbol{\phi}}) \rangle \nonumber\\
&\phantom{==} + \log Z_{\beta,\alpha}({\mathbf{Y}}) \nonumber\\
&= \beta U + \alpha V - \mathcal{G}(\beta,\alpha).
\label{eq:posterior_entropy}
\end{align}
The combination of Eq.~\ref{eq:partition_expectation} and Eq.~\ref{eq:posterior_entropy} forms a Legendre transformation that transforms $\mathcal{G}(\beta,\alpha)$ (a function of $\beta$ and $\alpha$) into the entropy $S(U,V)$ which is a function of $U$ and $V$.\footnote{In general, when a smooth convex function $f(x)$ is expressed in terms of a new function $f^{\ast}(p)=\max_{x}\{p x - f(x) \}$, this is called a Legendre transformation. However, since the derivative is zero at the maximum value, $p=f'(x)$, and thus $p$ represents the slope of the function $f(x)$. Hence, strictly speaking, Eq.~\ref{eq:posterior_entropy} multipled by $-1$ represents the Legendre transformation from $-\mathcal{G}(\beta,\alpha)$ to $-S(U,V)$}

Since there is no loss of information by the Legendre transformation defined by Eqs.~\ref{eq:partition_expectation} and \ref{eq:posterior_entropy}, it is always possible to return to the original function by using the inverse Legendre transformation. The inverse Legendre transformation is given by
\begin{equation}
\mathcal{G}(\beta,\alpha) = \beta U + \alpha V - S(U,V).
\label{eq:gibbs_freeenergy}
\end{equation}
and
\begin{equation}
\frac{{\partial S}}{{\partial U}} = \beta, \ \ \ \frac{{\partial S}}{{\partial V}} = \alpha.
\label{eq:entropy_partialderiv}
\end{equation}
This transformation converts the entropy $S(U,V)$ (a function of $U$ and $V$) into $\mathcal{G}(\beta,\alpha)$ (a function of $\beta$ and $\alpha$). When $\beta=1$ and $\alpha=1$, an exact posterior distribution is obtained, in which case the marginal likelihood and negative log partition function coincide: $\log p(\mathbf{Y} | \mathbf{w}) = - \mathcal{G}(1,1)$.\footnote{In general, the relationship with the lower bound is expressed as
\begin{equation}
\mathcal{L}[q,p] = -\mathcal{G}(\beta,\alpha) + (\beta - 1) U + (\alpha - 1) V.
\end{equation}
}

According to Eq.~\ref{eq:entropy_partialderiv}, the total derivative of the entropy of the recognition model, 
\begin{equation}
dS = \left( \frac{{\partial S}}{{\partial U}} \right)_{V} dU + \left( \frac{{\partial S}}{{\partial V}} \right)_{U} dV,
\end{equation}
can be expressed as follows:
\begin{equation}
dS = \beta dU + \alpha dV.
\label{eq:first_law}
\end{equation}
This formula represents the contribution of the prior distribution and the observation model (input stimuli) to the change of entropy in the recognition model, and dictates the law of conservation of entropy. In physics, Eq.~\ref{eq:first_law} is called the first law of thermodynamics (law of conservation of energy).\footnote{The first law of thermodynamics $TdS=dU + f dV$ is obtained from Eq.~\ref{eq:first_law}, using the temperature $T=1/\beta$ and force $f = \alpha / \beta$. In this case $dU$ is called the internal energy, which is a state variable. On the other hand, the terms $\dbar Q=TdS$ and $\dbar W=fdV$, which are called heat and work, are represented using a path-dependent incomplete derivative $\dbar$. Using these terms, the first law of thermodynamics is also written as $\dbar Q= dU + \dbar W$.} Since entropy is a state variable that is determined for a particular recognition model, the entropy change on the left side of Eq.~\ref{eq:first_law} is expressed as a difference in state variables at two close recognition models. However, the terms $\beta dU$ and $\alpha dV$ on the right side depend on the integration path since $\beta$ and $\alpha$ are functions of $U$ and $V$.

The parameters $\beta$ and $\alpha$ of the approximate posterior distribution represent contributions of the prior distribution and the likelihood function when constructing the recognition model. This recognition model becomes an exact posterior distribution when $\beta=1$ and $\alpha=1$. It is expected that there exists neural dynamics that progressively approaches the optimal state for the Bayesian inference. Thus it is important to consider the dynamics of $\beta$ and $\alpha$ when we hypothesize that the Bayesian inference is implemented by neural dynamics in the brain. 

For example, a likely scenario for the formation of a posterior distribution by a neural network is as follows: after nerve cells have fired in response to the presentation of a stimulus, the neural activity is modulated by feedback input (including information corresponding to the prior knowledge), whereby the observation and prior knowledge are fused. In \cite{shimazaki2015neurons,shimazaki2018neural}, the spontaneous firing of neurons corresponding to the prior distribution and the firing activity of neurons induced by a stimulus are represented by exponential family distributions, to which we apply the thermodynamic formulation introduced in this section. In particular, it was shown that, when stimulus response is modulated with a time delay due to feedback/recurrent input, neural dynamics forms an information-theoretic cycle (an analogue of heat engine, termed neural engine).\footnote{In these articles, thermodynamic analysis of a neural population was proposed by expressing spontaneous/background activity and stimulus-related activity of neurons by an exponential family distribution. In particular, if the level of the background activity is changed with a time delay due to feedback inputs, the response of neurons to a stimulus undergo gain control. The dynamics of such a delayed gain-control of the stimulus response turns out to be analogous to a heat engine of thermodynamics. This response cycle preserve information about stimuli with the presence of the delayed feedback signal, that would otherwise be lost.} Many studies have shown that the modulation of late components of stimulus response is related to attention, perceptual experience, short term memory, and subjective reward value \cite{libet1967responses,cauller1991neural,reynolds2000attention,super2001neural,sachidhanandam2013membrane,Manita2015,Schultz2016}. This approach allows us to quantify higher-order brain functions by measuring active portion of the computation as entropy emission related to the modulation of stimulus response.

\subsection{Learning and the principle of increasing entropy}
Several thermodynamic equations and the law of conservation of entropy have been concisely obtained by adopting the maximum entropy model as a recognition model. Let’s see what happens in the learning process of this model. In the discussion so far, the parameter $\mathbf{w}=\{{\boldsymbol{\phi}}, {\boldsymbol{\omega}}\}$ has been fixed without considering learning, although we discussed dynamics to construct the optimal recognition model by changing $\beta$ and $\alpha$. Let us admit that parameter $\mathbf{w}$ is also optimized by learning, and assume that the learning dynamics also follows the maximum entropy principle. That is, the learning plays a role of another factor besides $\beta$ and $\alpha$ (or $U$ and $V$) that change the entropy of the recognition model, and this factor always increase entropy. More specifically, the law of increasing entropy can be derived as a statistical law for causal dynamics in which the forward and backward processes are different, from the fluctuation theorem \cite{crooks1999entropy,seifert2012stochastic,ito2018unified}. Assuming such causal dynamics for learning, it is expected that
\begin{equation}
dS \geq \beta dU + \alpha dV.
\label{eq:second_law}
\end{equation}
This is equivalent to the second law of thermodynamics in physics, and it applies whenever an irreversible process takes place. Here we examine how such a causal learning rule relates to the optimization principles introduced in previous sections. 

Consider optimizing the parameters while keeping $\beta$ and $\alpha$ fixed. The free energy is a convenient quantity that can be used instead of entropy under such conditions. In fact, we will see that the quantity $\mathcal{G}(\beta,\alpha)$ is the free energy. The total derivative of $\mathcal{G}(\beta,\alpha)$ that is a function of $\beta$ and $\alpha$ becomes 
\begin{align}
d \mathcal{G}(\beta,\alpha)
&= d(\beta U + \alpha V) - dS \nonumber\\
&= (U d\beta + \beta dU) + (V d\alpha + \alpha dV) - dS \nonumber\\
&= U d\beta + V d\alpha,
\label{eq:dG}
\end{align}
where the first law (Eq.~\ref{eq:first_law}) is used at the last equality. It can also be obtained directly from the definition of total derivative
\begin{equation}
d \mathcal{G}(\beta,\alpha) = \left( \frac{{\partial \mathcal{G}}}{{\partial \beta}} \right)_{\alpha} d\beta + \left( \frac{{\partial \mathcal{G}}}{{\partial \alpha}} \right)_{\beta} d\alpha, 
\label{eq:dG_totalder}
\end{equation}
and from Eq.~\ref{eq:partition_expectation}. $\mathcal{G}(\beta,\alpha)$ is a thermodynamic quantity that takes $\beta$ and $\alpha$ as natural independent variables. Therefore it is particularly useful when these independent variables are fixed. In this article we call $\mathcal{G}(\beta,\alpha)$ the Gibbs free energy.\footnote{In thermodynamics, the dual function based on the Legendre transformation of internal energy $U$ is called thermodynamic potential (free energy). The internal energy is given by the formula $dU = TdS + f dV$, with $S$ and $V$ as natural independent variables. For example, the Helmholtz free energy is $F=U-TS$, and from the relationship $dF = d(U - TS) = dU - (dT S + T dS) = f dV - S dT$. Here we have $V$ and $S$ as natural independent variables. This is a Legendre transformation. At constant temperature, $dF = f dV$. Hence, the work done in an isothermal process can be expressed as a difference of Helmholtz free energy. Similarly, the Gibbs free energy in thermodynamics is defined as $G=F+fV$. This is convenient expression to use in isothermal and isobaric processes because $dG=dF-(df V + f dV)=-S dT + V df$. However, the Gibbs free energy $\mathcal{G}$ in this article is given by the relation $\mathcal{G}=\beta G$. When considering the recognition models of the brain and machine, the concepts of heat and work may aid understanding, but it is not clear whether they will actually bring direct benefits. Since the main focus of this article is on entropy, we have regarded the Legendre transformation of entropy (rather than internal energy) as free energy.} The reason why we derive the total derivative using Eq.~\ref{eq:dG} in stead of Eq.~\ref{eq:dG_totalder} is that it becomes clear that the following relationship holds when entropy is increased by learning:
\begin{equation}
d \mathcal{G}(\beta,\alpha) \leq U d\beta + V d\alpha,
\end{equation}
by applying Eq.~\ref{eq:second_law} to Eq.~\ref{eq:dG}. In particular, if causal learning occurs while $\beta$ and $\alpha$ are fixed, the Gibbs free energy decreases,
\begin{equation}
d \mathcal{G}(\beta,\alpha) \leq 0.
\end{equation}
That is, the following learning rule can be derived:
\begin{equation}
\frac{d\mathbf{w}}{dt} = - \epsilon \frac{\partial \mathcal{G}}{\partial \mathbf{w}},
\label{eq:learning_thermo}
\end{equation}
where $\mathbf{w}=\{{\boldsymbol{\phi}}, {\boldsymbol{\omega}}\}$, and $\epsilon$ is a learning coefficient of the parameters.\footnote{
Here we derived learning rules from the second law of thermodynamics or the minimization principle of free energy, but it is possible to decide on a specific causal learning rule and derive the law of increasing entropy (the second law of thermodynamics) \cite{goldt2017stochastic,salazar2017nonequilibrium}. This approach to the learning process started to be discussed with the recent development of stochastic thermodynamics \cite{ito2013information,hartich2014stochastic}. 
} 
Alternatively to Eq.~\ref{eq:learning_thermo}, an unique value of the learning coefficient may be considered for each parameter to account for adaptation to environmental dynamics with different time-scales.

From the above, when $\beta$ and $\alpha$ are fixed, the learning process that decreases the Gibbs free energy of the recognition model is equivalent to the learning that maximizes the entropy of the recognition model (the second law of thermodynamics). In particular, when $\beta=1$ and $\alpha=1$, the Gibbs free energy becomes a negative marginal likelihood, and learning according to the law of increasing entropy becomes equivalent to learning according to the marginal likelihood maximization. Therefore, if there is a mechanism for forming an optimal posterior distribution ($\beta=1$, $\alpha=1$) as a recognition model after a stimulus is received, and if learning is performed at this time, we can expect the results obtained by minimizing the Gibbs free energy are the same as those obtained by maximizing the marginal likelihood. One may also consider the hypothesis that actions are also selected so as to reduce the Gibbs free energy, following the assertions of Friston et al.

In this section, we clarified the relationship between learning according to the law of increasing entropy, learning according to the minimization of Gibbs free energy, and the maximization of marginal likelihood, by using a maximum entropy model to form a recognition model. In thermodynamics, free energy is introduced by finding a new thermodynamic quantity having the same meaning as the law of increasing entropy under certain conditions when we need to know in which direction a phenomenon will occur as prescribed by the second law of thermodynamics. Changes in gases and liquids at constant temperature and constant pressure take place only when some internal change such as a chemical reaction takes place, and such reactions proceed in a direction such that the Gibbs free energy decreases. We explained how learning can be treated in the same way in this section. 

\section{Summary and prospects}\label{sec:discussion}
In this article, we examined hypotheses on adaptive processes of organisms to their environments from multiple viewpoints: maximizing the marginal likelihood (maximizing the lower bound and minimizing the variational free energy), maximizing the mutual information, the law of increasing entropy (the second law of thermodynamics), and minimizing the free energy. 

One important topic which was not systematically covered in this article is the adaptation of organisms at different time scales. Data from the environment has a temporal hierarchy ranging from the formation of context over long timescales to short-term fluctuations. Organisms are equipped with multiple adaptive mechanisms for such environmental changes. For example, the intensity of light experienced by an organism changes with the cycle of day and night, but also undergoes sharp changes as the organism moves between light and shade. Animals perform multiple adaptation processes to cope with the intensity changes, starting with constriction of pupils known as the pupillary light reflex, and including relatively fast gain control performed in cells in retina \cite{Sakmann1969,shapley1978effect} and primary visual cortex \cite{ohzawa1982contrast}. Furthermroe, it was shown that visual attention of monkeys whereby the animals change sensitivity to light contrast according to the context of tasks is also explained by the gain control mechanism \cite{reynolds2000attention}. The attentional mechanisms in which organisms select data according to the context may also be explained by the canonical adaptation principle to the environment \cite{reynolds2000attention,salinas2001gain,reynolds2009normalization,carandini2012normalization,eldar2013effects}. This implies that slower temporal dynamics is required for the neural dynamics to retain contextual information occuring in a long time-scale. Indeed, it has been reported that the intrinsic time scale of neural activity slows down along the way from the primary visual cortex to the prefrontal cortex \cite{murray2014hierarchy}. Finally, the adaptation of visual stimuli to spatial structures is a process of adaptation to data distributions more slowly on a time scale corresponding to the organism’s own development. All of these adaptations are thought to be performed using biophysical phenomena operating on different time scales, including electrical responses and intracellular signal propagation (from tens of milliseconds to several seconds), and synaptic plasticity resulting from protein synthesis (from tens of minutes to several hours). It is necessary to clarify these hierarchical dynamics in order to understand the adaptation of organisms to the environment.\footnote{The method of extracting features based on differences in time scales is called slow feature analysis, and has been proposed as one of the brain’s guiding principles \cite{foldiak1991learning,wiskott2002slow,berkes2005slow}.}

We touched upon these topics on the temporal hierarchy in adaptation and learning at each section, but not in a systematic manner. In Section \ref{sec:learning} that introduced learning, we mentioned different sample size for learning parameters of the models. In addition to learning the parameters in the generative model from multiple samples for adaptation in a long time-scale, we also introduced a process to learn the prior distribution for each sample to account for the short-term adaptation. In the section of information theory (Section \ref{sec:information_theory}), using a simple example of the noiseless channel, we confirmed that the nonlinear function (neural network) should be adapted to the data distribution under the informax principle in both long and short time-scales. Further, we discussed how this nonlinear function can be mapped into the generative model, and saw that the optimal nonlinear function is modulated in the presence of the prior distribution. Finally, thermodynamic analysis in Section \ref{sec:thermodynamics} formulated relative contributions of the observation and prior to construct the recognition model, using the weight paramters $\beta$ and $\alpha$. In this framework, learning parameters of the observation model and prior distribution in a longer time-scale was achieved by minimization of the Gibbs free energy. Short-term dynamics of the adaptation was discussed as changes of the parameters $\beta$ and $\alpha$, which is described analogously to a thermodynamic process. Further, it was shown that this process works similarly to a heat engine when the stimulus response (observation) is modulated by feedback inputs (prior information) via top-down or lateral connections \cite{shimazaki2015neurons,shimazaki2018neural}. While the issues of adapting to different temporal scales have been discussed in other articles, a unified theory that can be brought into practice remains to be constructed.

It can be stated that organisms are equipped with a number of adaptive mechanisms to environments composed of spatial and temporal hierarchies by utilizing their biophysical phenomena with various time-scales, in order to maintain information-theoretic balance with the environment. With active inference that includes behavior, this balance is stabilized because the organisms build more predictable environments by use of the actions. By investigating these adaptive processes from multiple view points, we will keep gaining deeper understanding about their principles.

\subsection*{acknowledgement}
I would like to thank Manuel Baltieri, Seyed-Amin Moosavi, Sousuke Ito, Ryota Kobayashi, Shashwat Shukla, Masanori Murayama, and Takuma Tanaka for critical reading of the article and helpful discussions.

\bibliographystyle{ieeetr} 
\bibliography{references,statespaceising,neuralengine} 

\end{document}